\documentclass[runningheads,a4paper]{llncs}
\usepackage{graphicx}
\usepackage{amsmath}
\usepackage{amssymb}
\usepackage{xspace}

\usepackage{tikz}
\usetikzlibrary{decorations.pathmorphing,calc,shadows.blur,shadings}

\usetikzlibrary{arrows,automata}

\usetikzlibrary{shadows.blur}
\usetikzlibrary{shapes.symbols}

\usepackage{array}
\usepackage{verbatim}

\usepackage{stmaryrd}
\usepackage{amsmath}
\usepackage{amssymb}

\usepackage{arrows}

\usepackage[ruled]{algorithm2e}

\def\BLUE#1{\textcolor{blue}{#1}}

\usepackage{color}
\usepackage{xcolor}

\newcommand{\SET}[1]{\{#1\}}  
\newcommand{\aN}{^{(N)}}

\def\calU{{\cal U}}

\def\frc{\textsf{frc}\,}

\def\vt#1{#1}

\newcommand{\nats}{\mathrm{I\!N}}
\newcommand{\reals}{\mathrm{I\!R}}
\newcommand{\nnreals}{\reals_{\geq 0}}

\def\deriv{\noindent\hspace*{.20in}\vspace{0.1in}}

\def\BLUE#1{\textcolor{blue}{#1}}

\def\MAGENTA#1{\textcolor{magenta}{#1}}

\newcommand{\vr}[1]{\mathbf{#1}}

\newcommand\MN{M^{(N)}}
\newcommand\esp[1]{{\mathchoice{\besp{#1}}{\sesp{#1}}{\sesp{#1}}{\sesp{#1}}}}
\newcommand\besp[1]{\mathbb{E}\left[#1\right]}
\newcommand\sesp[1]{\mathbb{E}[#1]}

\newcommand\p[1]{\left(#1\right)}

\def\calU{{\cal U}}

\usepackage{doi}
\usepackage{url}
\newcommand\Gm{G_{\mathit{max}}}

\newcounter{mknot}
\newenvironment{mknot}[1][]{\refstepcounter{mknot}\par\medskip
   \noindent \textbf{\BLUE{NfM~\themknot.} #1} \rmfamily}{\medskip}

\newcounter{dgnot}
\newenvironment{dgnot}[1][]{\refstepcounter{dgnot}\par\medskip
   \noindent \textbf{\MAGENTA{NfD~\thedgnot.} #1} \rmfamily}{\medskip}

\providecommand{\url}[1]{{#1}}

\def\deriv{\noindent\hspace*{.20in}\vspace{0.1in}}

\usepackage{tabto}
\newcommand{\dertab}{.9cm}
\newcommand{\dervskip}{.2cm}

\newcommand{\der}{$\vskip \dervskip\noindent\tabto{\dertab}$}

\newcommand{\stp}[1]{\\*[\dervskip]#1$\tabto{\dertab}$}

\begin{document}

\mainmatter

%\begin{frontmatter}


%\title{Refined Mean Field Analysis of a Gossip Protocol}
\title{Refined Mean Field Analysis of the\\ Gossip Shuffle Protocol\\ -- extended version --}
%\titlerunning{Refined Mean Field Analysis of a Gossip Protocol}
\titlerunning{Refined Mean Field Gossip Shuffle Protocol}

\author{Nicolas~Gast\inst{1}\and Diego~Latella\inst{2} \and Mieke~Massink\inst{2}}
\institute{INRIA, France
\and
Consiglio Nazionale delle Ricerche - Istituto di Scienza e Tecnologie dell'Informazione \lq A.~Faedo\rq, CNR, Italy}

\authorrunning{Gast et al.}

\maketitle

\begin{abstract}
Gossip protocols form the basis of many smart collective adaptive systems. They are a class of fully decentralised, simple but robust protocols for the distribution of information throughout large scale networks with hundreds or thousands of nodes. Mean field analysis methods have made it possible to approximate and analyse performance aspects of such large scale protocols in an efficient way. Taking the gossip shuffle protocol as a benchmark, we evaluate a recently developed {\em refined} mean field approach. We illustrate the gain in accuracy this can provide for the analysis of medium size models analysing two key performance measures. We also show that refined mean field analysis requires special attention to correctly capture the coordination aspects of the gossip shuffle protocol. 
\end{abstract}


%\end{frontmatter}

% !TEX root =  gossip_rmf.tex

\section{Introduction and Related Work}

Many collective adaptive systems rely on the decentralised distribution of information. Gossip protocols (also known as epidemic or random walk protocols) have been proposed as a paradigm that can provide a stable and reliable method for such decentralised spreading of information~\cite{VoulgarisJS03,Birman07,BakhshiGFS09,Jelasity11,FreiS11,FreiS11a,BakhshiGFS11,BakhshiCFH11,PianiniBV16}. Gossip protocols are able to scale up to the very large environments that collective adaptive systems are envisioned for. The basic mechanism of information spreading followed by a gossip protocol is that nodes exchange part of the data they keep in their cache with randomly selected peers in pairwise synchronous communications on a regular basis.

Interesting performance aspects of such gossip protocols are the diffusion or replication of a newly inserted fresh data element in a network and the dynamics of network coverage. Diffusion or replication of a data element occurs when nodes exchange the data element in pairwise communication. Two relevant measures are of interest in this case. One is the fraction of the population that has the data element in its cache at a certain point in time (replication). The other concerns network coverage (coverage), i.e. the fraction of the population of network nodes that have ``seen'' the data element since its introduction into the network, even if they may no longer have it in their cache due to further exchanges with other peers. 

Traditionally, these performance measures have been studied based on simulation models. However, when large populations of nodes are involved, such simulations may be very resource consuming. Recently these protocols have been studied using classic mean field approximation techniques~\cite{BakhshiCFH11,Bak11}. In that classic approach the full stochastic model of a gossip network, i.e. one in which each node is modelled individually, is replaced by a much simpler model in which the pairwise synchronous interactions between individual nodes are replaced by the average effect that all those interactions have on a single node and then the model of this single node is studied in the context of the overall average network behaviour. Of course, the average effects may change over time as nodes may change their local states. This is taken into account in a mean field model by letting the probabilities of interactions possibly depend on the fraction of nodes that are in a particular local state. 
Compared to traditional simulation methods, mean field approximation techniques scale very well to large populations because these techniques are independent of the exact population size\footnote{As long as this size is large enough to obtain a sufficiently accurate approximation. The computational complexity of these techniques {\em does} depend on the number of local states of an object in a population.}. This method of derivation of a mean field model from a large population of interacting objects relies on what is known as the assumption of ``propagation of chaos'' (also called ``statistical independence'' or ``decoupling of joint probabilities'')~\cite{BMM07qest,BHLM13peva,gastgaujalDEDS,latella2013fly}. The assumption is based on the fact that when the number of interacting nodes becomes very large, their interactions tend to behave as if they were statistically independent.

\begin{figure}[h] 
   \centering
    \includegraphics[width=2.3in]{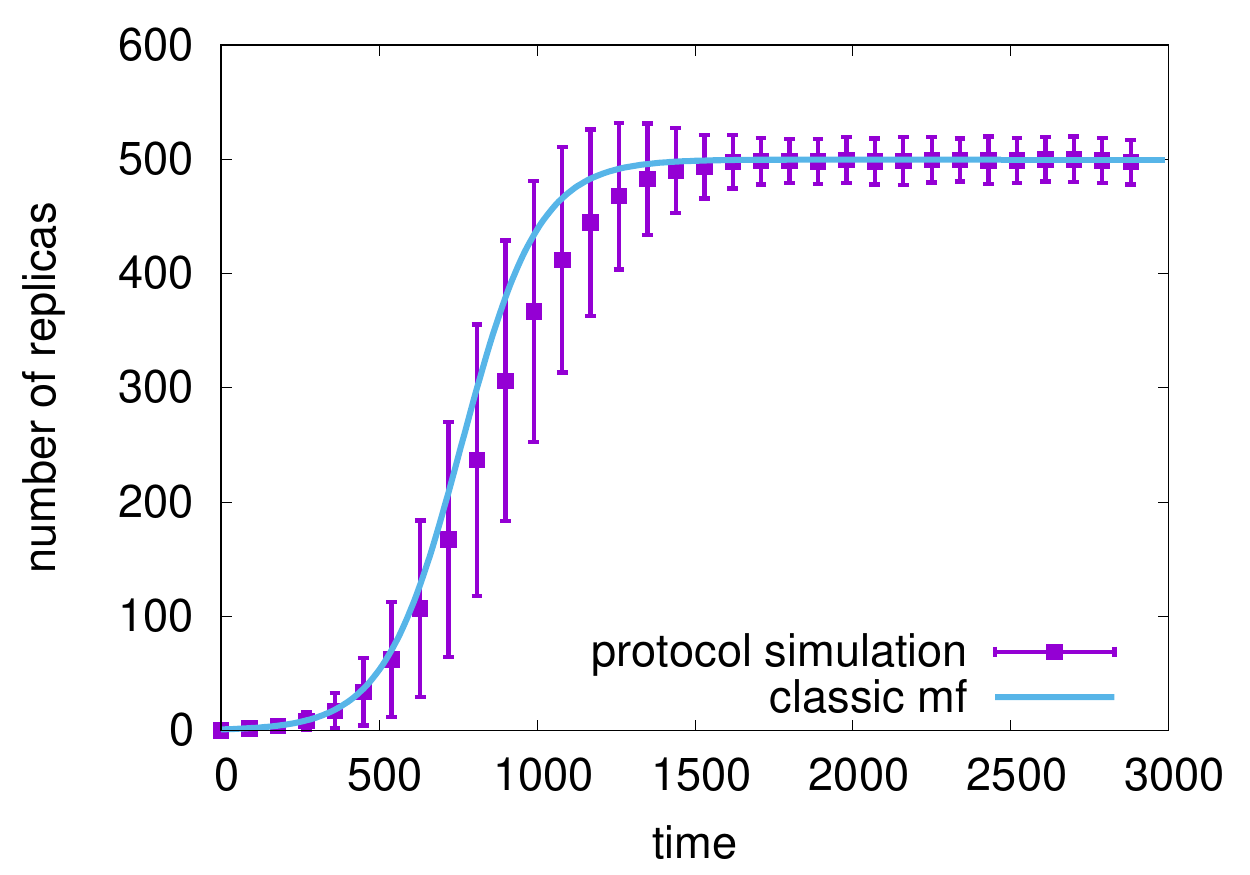} 
    \includegraphics[width=2.3in]{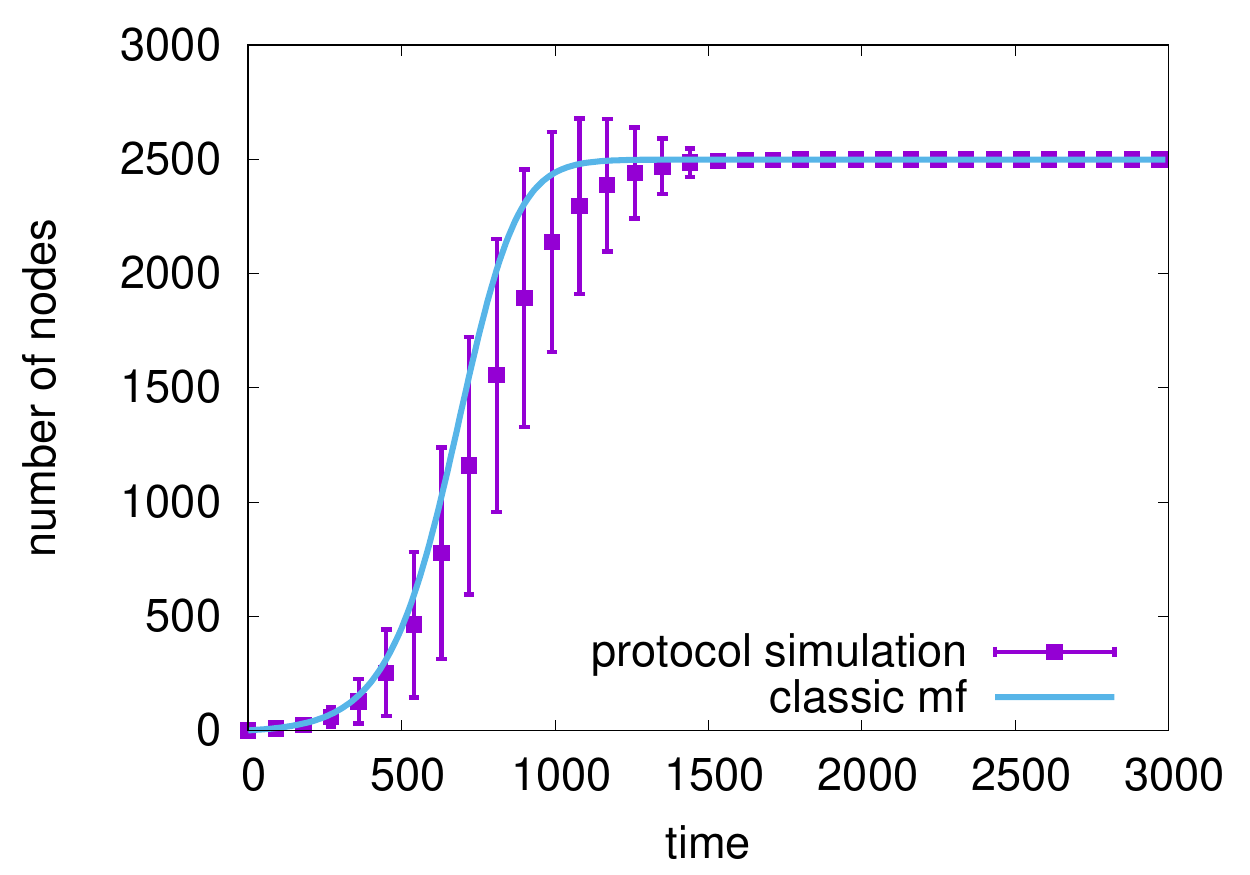} 
\caption{\label{fig:gossip3statesAgg}Replication (left) and Coverage (right) for one new data element in a network with $N=2500$.
% and with $n=500$ different data elements in the network where initially 1 node possesses the new data element and the maximal time that passes before an active node gets active again is  $\Gm=9$. 
 Average of 500 simulation runs of the Java simulator~\cite{Bak11}. Vertical bars show standard deviation for the simulation. }
\end{figure}

However, in reality, we are not always dealing with huge systems, but rather with medium size ones. These are still resource intensive when analysed using simulation and, unfortunately, the classical mean field approximation is less accurate for such medium size systems. For example, in Fig.~\ref{fig:gossip3statesAgg} the results of classical mean field approximation are shown together with a Java based simulation of the protocol for a medium size gossip system with 2500 nodes where initially one node has a new data element that will spread over the network by gossiping.
It is easy to see that there is a discrepancy between simulation and classic mean field approximation, both for replication of the data element and for the coverage, even in this not so small system. 

In this paper we revisit an analysis of the gossip shuffle protocol using a {\em refined mean field approximation} for {\em discrete time population models} that we developed in~\cite{GLM18,GLM2019}, and which was in turn inspired by an earlier result for continuous time population models in~\cite{GastH17}.
The gossip shuffle protocol was analysed in detail by Bahkshi et al. in~\cite{BakhshiGFS11,BakhshiCFH11,Bak11} both analytically and by using classic mean field approximation in~\cite{BakhshiCFH11,Bak11} and, more recently, by using on-the-fly mean field discrete time model checking techniques in~\cite{LatellaLM17}. The present paper is an extended version of the short paper~\cite{GLM2020a}.

\paragraph*{Contributions}
The main contribution of this paper is a novel benchmark (clock-synchronous) DTMC population model of the gossip shuffle protocol analysed using our refined mean field analysis~\cite{GLM18,GLM2019}. % as developed in earlier work by the authors~\cite{GLM18,GLM2019}. 
In particular:
\begin{itemize}
\item We show that with refined mean field approximation better accuracy can be obtained compared to classical mean field approximation for medium size populations for this gossip protocol, but that this requires a novel model that reflects the synchronisation effects of the pairwise interaction of the original protocol.  
\item The developed model is parametric in $G_{max}$, i.e. the number of steps it remains passive in between active interactions with peer gossip nodes.
\item The results we obtained are very close  both to those of independent Java based simulation from the literature in~\cite{BakhshiCFH11} (taken as ``ground truth'') and to those of the event simulation of the model itself, but with the advantage that the refined mean field approximation is several orders of magnitude faster to obtain and {\em independent} of the system size. 
\item Development of a proof-of-concept implementation in {F$\#$} of both the classical and the refined mean field techniques and a discrete event simulator used for the analysis of the gossip shuffle protocol~\cite{Mas2020}.
\end{itemize}

Like classic mean field approaches, the refined approach is computationally non-intensive and {\em the analysis time is independent of the population size}. The analysis is orders of magnitudes faster than discrete event based simulation. Therefore it is an interesting candidate for being integrated with other analysis approaches such as (on-the-fly) mean field model checking, which is planned in future work. The current study aims at providing further insight in the feasibility of applying the refined mean field approach, that implies the use of symbolic differentiation, on larger benchmark examples and in the possible complications of such an analysis that need to be taken into consideration.

The outline of the paper is as follows. The relevant aspects of the gossip shuffle protocol are briefly recalled in Section~\ref{sec:gossip}.
The refined mean field approach used in this paper applies to the classical population model of~\cite{BMM07qest,gastgaujalDEDS,latella2013fly} and is briefly recalled in Section~\ref{sec:background}. Section~\ref{sec:gossipCMF} presents full and aggregated classical mean field models of the protocol which form the starting point for the novel gossip model suitable for refined mean field approximation presented and analysed in Section~\ref{sec:gossipRMF}. Section~\ref{sec:concl} presents conclusions.

% !TEX root =  gossip_rmf.tex

\section{Benchmark Gossip Shuffle Protocol}
\label{sec:gossip}

We briefly recall the main aspects of the gossip shuffle protocol described in~\cite{GVS2006,Bak11,BakhshiCFH11} that serves as our benchmark. This particular version has been extensively studied by Bahkshi et al., leading to an analytical model of the gossip protocol~\cite{BakhshiGFS09}, a classical mean field model~\cite{BakhshiCFH11} and a Java implementation of a simulator for the protocol~\cite{BakhshiCFH11}, which makes it a very suitable candidate of a real-world application that allows for the comparison of new results. In the following we briefly recall some main aspects of the shuffle gossip protocol and the Java simulator. Further details can be found in~\cite{BakhshiCFH11,Bak11}.

%This protocol is extensively studied and described in detail in~\cite{GVS2006,Bak11,BakhshiCFH11}, which  with earlier ones in the literature. Here we only recall some of the main aspects. In particular, 

\subsection{Informal description}
The gossip shuffle protocol distributes data items throughout a network of small devices. Such networks typically consist of a very large collection of nodes. Each node has a limited amount of storage space (called its cache) for the data items. At any instant, gossip nodes are divided into two classes: active and passive nodes. Active nodes can initiate a shuffle, i.e. an exchange of data between two peers, by contacting a passive neighbouring node and exchange part of their data. Such a passive node is selected through an underlying layer\footnote{This layer is not explicitly modelled. For example, in wireless environments such passive peers may be determined by the radio connectivity between nodes.} that keeps track of which nodes are active or passive.

\begin{figure}
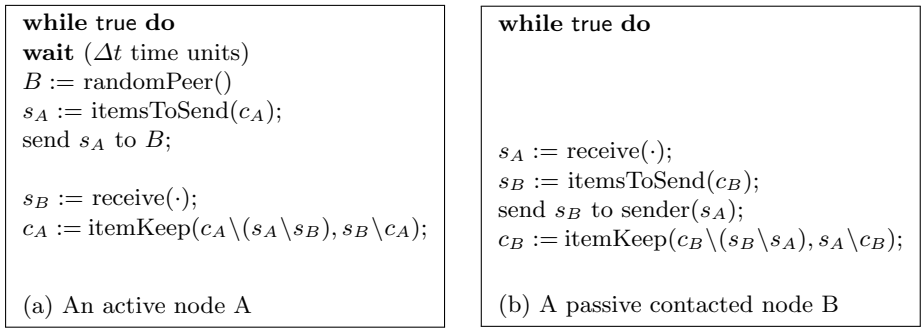

\begin{center}
\fbox{
\begin{minipage}{2.1in}
{\bf while} {\sf true} {\bf do}\\
{\bf wait} ($\Delta t$ time units)\\
$B :=$ randomPeer()\\
$s_A :=$ itemsToSend($c_A$);\\
send $s_A$ to $B$;\\\\
$s_B :=$ receive($\cdot$);\\
$c_A :=$ itemKeep($c_A\setminus (s_A\setminus s_B),s_B\setminus c_A$); \\[2em]
(a) An active node A
\end{minipage}
}\;\;\;\;
\fbox{
\begin{minipage}{2.1in}
{\bf while} {\sf true} {\bf do}\\[4em]
$s_A :=$ receive($\cdot$);\\
$s_B :=$ itemsToSend($c_B$);\\
send $s_B$ to sender($s_A$);\\
$c_B :=$ itemKeep($c_B\setminus (s_B\setminus s_A),s_A\setminus c_B$);\\[1.5em]
(b) A passive contacted node B
\end{minipage}
}
\end{center}
\caption{\label{fig:gossip} Pseudo code of a generic shuffle protocol (adapted from ~\cite{Bak11}). $c_A$ and $s_A$ denote the cache and selection of active node $A$. Similarly, $c_B$ and $s_B$ denote those of passive node $B$. $\Delta t= \Gm$. The operation `itemsToSend($c_i$)' selects the items to be sent from the cache $c_i$. The operation `itemKeep(c,s)' in node A decides which items to keep in the cache (c) removing from the cache those selected for sending ($s_A$) except those that where received from B ($s_B$), and adding to those the elements from $s_B$ that were not yet in the cache of A. Similarly for the operation in node B.}
\end{figure}

Each gossip node maintains a finite list of data items in its cache. Both the active node and its passive partner exchange a random subset from their local caches in one atomic peer-to-peer communication session. Given the limited size of the cache, a node may have to discard some items it receives. This is done in such a way that {\em no information is lost in the network}, i.e. a node discards items selected among those that it has just sent to its peer and does not discard new items it has just received from the peer.
Fig.~\ref{fig:gossip} recalls the pseudo code of a generic shuffle protocol (adapted from~\cite{Bak11}). 

% See also page 21 Bakhshi thesis for a nicer illustration. Maybe we can replace the above?
Two main key measures that are of interest for this protocol are the transient aspects of the {\em replication} of a newly introduced element in the network and that of the {\em coverage} of the network, i.e. the fraction of network nodes that have seen the new data element when time is passing. These measures depend on a number of characteristics of the network. In the following we use $N$ to denote the size of the network, i.e. the number of gossiping nodes, $n$ to denote the number of {\em different} data items in the network, $c$ to denote the size of the cache and $s$ to denote the size of the selected items from the cache to be exchanged with a neighbour. In the context of this work, and for comparison with the results presented in~\cite{Bak11}, the network is assumed to be {\em fully connected}. We consider a discrete time variant of the protocol with a maximal delay between two subsequent active data-exchanges of a node denoted by $\Gm$. 

\subsection{The gossip Java simulator}
To assess the quality of classic mean field approximation results, Bahkshi et al. developed a Java-based implementation of a simulator for the shuffle protocol with which networks of various sizes can be simulated on a single processor~\cite{BakhshiCFH11}. In this paper we also adopt the results produced by this simulator, the source code of which was generously shared with us by the developers, as the ``ground truth" with which to compare our own results. This simulator works as follows. It takes the network size $N$, and the specific size of the storage, $c$, the number of messages exchanged in each shuffle, $s$ and the total number of different data elements in the network, $n$. It divides all network nodes into $\Gm+1$ different groups, each representing a different value of the gossip delay. Recall that the maximal period between two consecutive contact initiations of any particular network node is $\Gm$. The nodes in the group with gossip delay equal to zero are the active nodes, i.e. those that initiate contact with their peers in the current round uniformly at random. If an active node contacts a node that is already in contact with another node, the interaction between all three nodes fails, leading to a collision. At the start of the simulation, a new data item is introduced in the network (i.e. one different from the $n$ types of data-elements that are already present in the network and that are assumed to be uniformly distributed over the local cash of all network nodes. After each round, the total number of copies of the new data element in the network  (replication) and the number of nodes that have seen the data element (coverage) are measured.

% !TEX root =  gossip_rmf.tex

\section{Background}
\label{sec:background}

In the sequel we use theoretical results on discrete time mean field approximation \cite{BMM07qest,BHLM13peva,GLM18}. We briefly recall the notation and main results in the following.
We consider a population model of a system composed of $0 < N \in \nats$ identical interacting objects, i.e.
a (model of a) system of {\em size} $N$.  
%It is worth noting that this characterisation includes
%also a system with {\em different} classes of {\em identical} objects; the latter can easily 
%be modelled by considering an equivalent system with instances of an object whose set of states is
%the union of those of the original objects and similarly for the set of its transitions.
%In this paper, 
We assume  that the set $\SET{0,\ldots, n-1}$ of local states
of each object is finite; we refer to~\cite{GLM18} for a discussion on how 
to deal with infinite dimensional models.  Time is {\em discrete} and the behaviour of
the system is characterised by a (time homogeneous) {\em discrete time
Markov chain} (DTMC) $\vt{X}\aN(t)=(X_1\aN(t), \ldots, X_N\aN(t))$,
where $X_i\aN(t)$ is the state of object $i$ at time $t$, for $i=1,\ldots, N$.

The {\em occupancy measure vector} at time $t$ of the model is the row-vector DTMC
$\vt{M}\aN(t)=(M_0\aN(t), \ldots, M_{n-1}\aN(t))$ where, for
$j=0,\ldots, n-1$, the stochastic variable $M_j\aN(t)$ denotes the {\em fraction} of objects in state
$j$ at time $t$, over the total population of $N$ objects:
\begin{align*}
  M_j\aN(t) = \frac{1}{N}\sum_{i=1}^N 1_{\SET{X_i\aN(t)=j}}
\end{align*}
and $1_{\SET{x=j}}$ is equal to $1$ if $x=j$ and $0$ otherwise.
At each time step $t \in \nats$ each object performs a local
transition, possibly changing its state.  The transitions of any two objects 
are assumed to be independent from each other, while the
transition probabilities of an object  may depend also on $M(t)$, thus, for large $N$,
the probabilistic behaviour of an object is characterised by the 
one-step transition probability $n \times n$ matrix $\vr{K}(\vt{m})$, where
$\vr{K}_{ij}(\vt{m})$
is the probability for the object to jump from state $i$ to state $j$ when the occupancy measure vector is $m \in \calU^n$;  $\calU^n$ is the unit simplex of $\nnreals^n$, that is
$\calU^n=\SET{\vt{m} \in [0,1]^n \;|\; \sum_{i=1}^{n} m_i =1}$.
%$\calU^n\defeq \Set{(m_0, \ldots , m_{n-1} \in \[0,1\]^n \mid m_0+\cdots + m_n=1}$
%where $\calU^2$ is the unit simplex of $\nnreals^2$. 
In this paper, for simplicity, we
assume $\vr{K}(\vt{m})$ to be a continuous function of $\vt{m}$ that does not depend on $N$.
%
%In the sequel, for the sake of readability, we will often use symbolic names 
%$\mathtt{C},  \mathtt{C}_1, \ldots, \mathtt{C}_r$  for states, and similarly
%$\mathtt{a}, \mathtt{a}_1, \ldots, \mathtt{a}_r $ for actions identifying transitions; 
%the transitions out of a state
%$\mathtt{C}$ will be defined symbolically using the following syntax:
%$\state \; \mathtt{C} \; \SET{\mathtt{a}_1 . \mathtt{C}_1 + \ldots + \mathtt{a}_r . \mathtt{C}_r}$,
%the intended meaning being that, when in state $\mathtt{C}$,
%the object can jump to state $\mathtt{C}_1$, by firing a transition labelled by action $\mathtt{a}_1$,
%or to state $\mathtt{C}_2$, by firing a transition labelled by action $\mathtt{a}_2$, and so on.
%The probability assigned to the transition labelled by action $\mathtt{a}$ is declared
%by means of  an action probability definition of the form 
%$\action \; \mathtt{a}: exp$ where $exp$ is an expression involving  $\frc(\mathtt{C})$ terms, the latter denoting the fraction of the objects currently in state $\mathtt{C}$, i.e. the element of the current occupancy measure vector
%associated to state $\mathtt{C}$. 
In the sequel, for reasons of presentation, we provide a graphical specification of the relevant models.
The computation of matrix $\vr{K}(\vt{m})$ from 
such a model specification is straightforward.

\subsection{Discrete Time Classical Mean Field Approximation}
Below we recall Theorem 4.1 of~\cite{BMM07qest} on classic mean field
approximation, under the simplifying assumptions mentioned above:

\begin{quotation}
\noindent
{\bf Theorem 4.1 of \cite{BMM07qest} (Convergence to Mean Field)}
{\em Assume that the initial occupancy measure $\vt{M}\aN(0)$ converges almost surely to the deterministic limit $\vt{\mu}(0)$. 
Define $\vt{\mu}(t)$ iteratively by (for $t \geq 0$):
\begin{align}
  \label{eq:mu}
  \vt{\mu}(t+1) = \vt{\mu}(t) \, \vr{K}(\vt{\mu}(t)).
\end{align}
Then for any fixed time $t$, almost surely, 
$
\lim_{N \to \infty} \vt{M}\aN(t) = \vt{\mu}(t).
$
}
\end{quotation}
%In the sequel, we will write $\vt{M}(t)$ or simply $\vt{M}$ instead of
%$\vt{M}\aN(t)$, leaving $N$ and $t$ implicit, when this does not cause
%confusion.

The above result thus allows one to use, for  {\em large} $N$,  a {\em deterministic} approximation $\vt{\mu}$ of the
average behaviour of a discrete population model.

% !TEX root =  gossip_rmf.tex

\subsection{Discrete Time Refined Mean Field Approximation}
\label{sec:refined_mf}

In~\cite{GLM18} we proposed a refined mean field method for {\em discrete time population models} that has shown to provide a considerably better approximation than classic mean field in the case of population models with a {\em medium population size} $N$. This work was inspired by the development of a refined mean field approximation for continuous time population models in~\cite{GastH17}.  Before recalling the theoretical results for the refined mean field approximation technique for discrete time models we introduce some further basic notation.

$\nnreals^n$ denotes the set of $n$-tuples---i.e. $1 \times n$  matrices---of non-negative real numbers.
For $n \times m$ matrix  $A$ we let $A^T$ denote its $m \times n$ {\em transposed} matrix.
For function $f :\reals^n \rightarrow \reals^p$ continuous and twice differentiable, %with $ f(\vt{m}) = (f_1(\vt{m}), \ldots, f_p(\vt{m})), $ we 
let 
the $p \times n$ (function) matrix $Df(\vt{m})$ and 
the $p \times n \times n$ tensor  $D^2f(\vt{m})$ denote its first and second derivatives, respectively:
$(D f (\vt{m}))_{ij} = \frac{\partial f_i(\vt{m})}{\partial m_j}$ and
$(D^2 f (\vt{m}))_{ijk} = \frac{\partial^2 f_i(\vt{m})}{\partial m_j\partial m_k}$.
Let function $\Phi: \nats \to \calU^n \to \calU^n$ % $\Phi_{t}(\vt{m})$ of $t$ and $m$ 
be defined as follows:
\begin{align*}
  \Phi_{0}(\vt{m}) = \vt{m}; \qquad {\displaystyle (\Phi_1(\vt{m}))_j =
  \sum_{i=0}^{n-1} m_i \vr{K}_{ij}(\vt{m})};
  \qquad \Phi_{t+1}(\vt{m}) = \Phi_1(\Phi_t(\vt{m})).
\end{align*}

Note that $\Phi_1(\Phi_t(\vt{m}))= \Phi_t(\Phi_1(\vt{m}))$ and that, for $\mu(t)$ defined as in Equation (\ref{eq:mu}), we have:
$\mu(t+1)=\Phi_1(\mu(t))=\Phi_{t+1}(\mu(0))$; so, function $\Phi$ makes explicit the 
dependence of $\mu(t)$ on the initial occupancy measure vector $m$.
Suppose function $h: \calU^n \to \nnreals^p$ models a {\em measure of interest} over the
occupancy measure vectors.  

Below we recall Theorem 1 we proved in~\cite{GLM18} on {\em Refined mean-field
approximation}:

\begin{quotation}
\label{theorem1}
\noindent
{\bf Theorem 1 of \cite{GLM18} (Refined Mean Field)}
{\em 
 Assume that function $\Phi_1$ is twice differentiable with
  continuous second derivative and that $M\aN(0)$ converges weakly to
  $\mu(0)$. Let $A_t$ and $B_t$ be respectively the $n \times n$
  matrix $A_t = (D \Phi_1)(\mu(t))$ and the $n \times n \times n$
  tensor $B_t = (D^2 \Phi_1)(\mu(t))$.  Then for any continuous and
  twice differentiable function with continuous second derivative
  $h:\calU^n \rightarrow \nnreals^p$ we have:
$$
\lim_{N\rightarrow \infty} N\esp{h(\vt{M}\aN(t))- h(\Phi_t(\vt{M}\aN(0)))} =
Dh(\mu(t)) V_t + \frac{1}{2}D^2h(\mu(t))\cdot W_t,
$$
where $V_t$ is an $n \times 1$ vector and $W_t$ is an $n \times n$ matrix, defined as follows:
$$
\begin{array}{lcl c lcl}
V_{t+1} & = & A_tV_t + \frac{1}{2}B_t \cdot W_t & \mbox{and} & %\\\\
W_{t+1} & = & \Gamma(\mu(t)) + A_t W_t A_t^T,
\end{array}
$$
with $V_0=0$, $W_0 = 0$ and $\Gamma(\vt{m})$ is the following
$n \times n$ matrix:
$$
\begin{array}{lcl}
\Gamma_{jj}(\vt{m}) & = &  \sum_{i=0}^{n-1} m_i \vr{K}_{ij}(\vt{m})(1-\vr{K}_{ij}(\vt{m}))\\\\
\Gamma_{jk}(\vt{m}) & = & -\sum_{i=0}^{n-1} m_i \vr{K}_{ij}(\vt{m})\vr{K}_{ik}(\vt{m})
\end{array}
$$
}
\end{quotation}

The following corollary illustrates the relationship between the refined mean field
result and the classic convergence theorem:

\begin{quotation}
\noindent
{\bf Corollary 1(i) of \cite{GLM18}}
{\em 
  Under the assumptions of {\bf Theorem 1 of \cite{GLM18}},
 it holds that for any coordinate $i$ and any time-step $t \in \nats$
    \begin{align*}
      \esp{\MN_i(t)} = \mu_i(t) + \frac{(V_{t})_i}{N} +  o\p{\frac1N}.
    \end{align*}
}
\end{quotation}

In other words, the expected value of the fraction of the objects in local state $i$ of the full stochastic model with population size $N$ at time $t$, is equal to the classic limit mean field value $\mu_i(t)$ plus a factor that is a constant $(V_{t})_i$, calculated as shown in Theorem 1, divided by the population size $N$ plus a residual amount of order $o\p{\frac1N}$. It is easy to see that the larger is N the smaller this additional factor gets. Essentially, the refined mean field takes not only the first moment (the mean) but also the second moment (variance) into consideration in the approximation.

In~\cite{GLM18} we have applied this discrete time refined mean field approximation on a number of examples ranging from the well-known epidemic model SEIR to wireless networks. It was shown that the approach works well under the assumption that the models have a unique fixed point and {\em exponentially stable behaviour}, i.e. possible oscillations in the behaviour of the system, due to a finite input, will die out at an exponential rate. Here we investigate its application to the more complex gossip shuffle protocol.

%A further result that can be obtained in {\em analogy} to that for the continuous time case in~\cite{GastH17} (Corollary 1) is that from the  $n \times n$ matrix $W_t$ the co-variance of random variables in the model can be obtained. In particular:
%
%\begin{quotation}
%\noindent
%{\bf Corollary 1(ii) of \cite{GastH17}}
%{\em 
%  For any coordinates i, j, the covariance satisfies:
%  $$
%  \lim_{N\rightarrow \infty} Ncov (M_i^{(N)},M_j^{(N)}) = W_{ij}.
%  $$
%}
%\end{quotation}
%
%This means that we can also obtain the standard deviation for the occupancy measure for any local state $i$ as $\sqrt{Ncov(M_i^{(N)},M_i^{(N)})} = \sqrt{W_{ii}} + O(1/\sqrt{N})$.

 A proof-of-concept implementation of both the classical and the refined mean field techniques and a discrete event simulator has been developed by one of the authors of the present paper in %the functional programming language
F${\#}$ using the DiffSharp package~\cite{BaydinPRS18} for symbolic differentiation. The results in this paper have been obtained using this implementation which can be found at~\cite{Mas2020}.

% and illustrate potential problems due to the stochastic independence assumption. 

% !TEX root =  gossip_rmf.tex

\section{Gossip Shuffling Protocol Mean Field Model}
\label{sec:gossipCMF}

Following the classic discrete time mean field approximation technique~\cite{BMM07qest,Bak11,BakhshiCFH11} the behaviour of an individual gossip node can be described based on its local state and the current occupancy measure vector. This exploits what is known as the ``decoupling principle", i.e. in the limit for $N$ going to infinity, the evolution of each individual object is assumed to be {\em stochastically independent} from other specific objects -- except through dependence on the global occupancy measure -- even in the presence of explicit cooperation (i.e. synchronisation) between objects~\cite{BMM07qest,BHLM13peva,Got13}. %Le+07 page 6 prop chaos, Got13 Gottlieb page 31 Chaos on finite sets
Such a model of an individual node can then be used to analyse global properties of the network such as the replication and coverage measures that are relevant in this case study. 

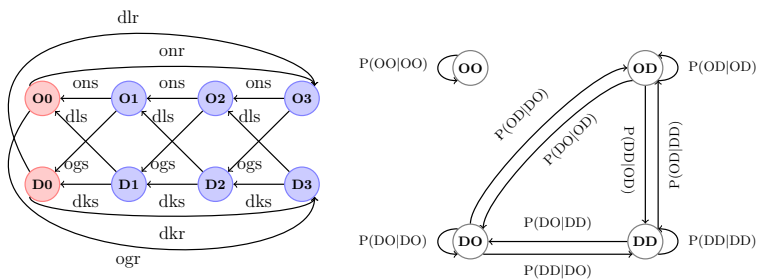
\begin{figure}
\begin{center}
\resizebox{0.45\textwidth}{!}{
%\raisebox{10em}{
\begin{tikzpicture}
\tikzstyle{place}=[circle,draw=blue!50,fill=blue!20,thick,inner sep=0pt,minimum size=8mm]
% Nodes with data-element
\node (D0) [place,draw=red!50,fill=red!20] at (0,0) {$\mathbf{D0}$};
\node (D1) [place,draw=blue!50,fill=blue!20] at (2,0) {$\mathbf{D1}$};
\node (D2) [place,draw=blue!50,fill=blue!20] at (4,0) {$\mathbf{D2}$};
\node (D3) [place,draw=blue!50,fill=blue!20] at (6,0) {$\mathbf{D3}$};

% Nodes without data-element
\node (O0) [place,draw=red!50,fill=red!20] at (0,2) {$\mathbf{O0}$};s
\node (O1) [place,draw=blue!50,fill=blue!20] at (2,2) {$\mathbf{O1}$};
\node (O2) [place,draw=blue!50,fill=blue!20] at (4,2) {$\mathbf{O2}$};
\node (O3) [place,draw=blue!50,fill=blue!20] at (6,2) {$\mathbf{O3}$};

%\draw[->,thick] (S.west) to [out=180,in=90] (S.north);
\draw[->,thick] (D3.west) .. controls +(up:0mm) and +(up:0mm) .. (D2.east) node[pos=0.5, label=below:{{\large dks}}]{};
\draw[->,thick] (D2.west) .. controls +(up:0mm) and +(up:0mm) .. (D1.east) node[pos=0.5, label=below:{{\large dks}}]{};
\draw[->,thick] (D1.west) .. controls +(up:0mm) and +(up:0mm) .. (D0.east) node[pos=0.5, label=below:{{\large dks}}]{};

\draw[->,thick] (O3.west) .. controls +(up:0mm) and +(up:0mm) .. (O2.east) node[pos=0.5, label={{\large ons}}]{};
\draw[->,thick] (O2.west) .. controls +(up:0mm) and +(up:0mm) .. (O1.east) node[pos=0.5, label={{\large ons}}]{};
\draw[->,thick] (O1.west) .. controls +(up:0mm) and +(up:0mm) .. (O0.east) node[pos=0.5, label={{\large ons}}]{};

\draw[->,thick] (D3.north west) .. controls +(up:0mm) and +(up:0mm) .. (O2.south east) node[pos=0.6, label={{\large dls}}]{};
\draw[->,thick] (D2.north west) .. controls +(up:0mm) and +(up:0mm) .. (O1.south east) node[pos=0.6, label={{\large dls}}]{};
\draw[->,thick] (D1.north west) .. controls +(up:0mm) and +(up:0mm) .. (O0.south east) node[pos=0.6, label={{\large dls}}]{};

\draw[->,thick] (O3.south west) .. controls +(up:0mm) and +(up:0mm) .. (D2.north east) node[pos=0.6, label=below:{{\large ogs}}]{};
\draw[->,thick] (O2.south west) .. controls +(up:0mm) and +(up:0mm) .. (D1.north east) node[pos=0.6, label=below:{{\large ogs}}]{};
\draw[->,thick] (O1.south west) .. controls +(up:0mm) and +(up:0mm) .. (D0.north east) node[pos=0.6, label=below:{{\large ogs}}]{};

\draw[->,thick] (O0.north west) .. controls +(up:6mm) and +(up:6mm) .. (O3.north east) node[pos=0.5, label={{\large onr}}]{};
\draw[->,thick] (D0.south west) .. controls +(down:6mm) and +(down:6mm) .. (D3.south east) node[pos=0.5, label=below:{{\large dkr}}]{};

\draw[->,thick] (O0.south west) .. controls (-3,-2)  and +(down:20mm) .. (D3.south east) node[pos=0.5, label=below:{{\large ogr}}]{};
\draw[->,thick] (D0.north west) .. controls (-3,5) and +(up:12mm) .. (O3.north east) node[pos=0.5, label={{\large dlr}}]{};

\end{tikzpicture}
}\quad
\resizebox{0.45\textwidth}{!}{
\begin{tikzpicture}
\tikzstyle{place}=[circle,draw=white!50,fill=white!20,thick,inner sep=0pt,minimum size=8mm]
% Nodes with data-element
\node (OO) [place,draw=black!50,fill=white!20] at (0,4) {$\mathbf{OO}$};
\node (OD) [place,draw=black!50,fill=white!20] at (4,4) {$\mathbf{OD}$};
\node (DD) [place,draw=black!50,fill=white!20] at (4,0) {$\mathbf{DD}$};
\node (DO) [place,draw=black!50,fill=white!20] at (0,0) {$\mathbf{DO}$};

\draw[->,thick] (OO.north west) .. controls +(left:6mm) and +(left:6mm) .. (OO.south west) node[pos=0.4, label=left:{{P(OO$|$OO)}}]{};
\draw[->,thick] (DO.north west) .. controls +(left:6mm) and +(left:6mm) .. (DO.south west) node[pos=0.5, label=left:{{P(DO$|$DO)}}]{};
\draw[->,thick] (DD.south east) .. controls +(right:6mm) and +(right:6mm) .. (DD.north east) node[pos=0.5, label=right:{{P(DD$|$DD)}}]{};
\draw[->,thick] (OD.south east) .. controls +(right:6mm) and +(right:6mm) .. (OD.north east) node[pos=0.5, label=right:{{P(OD$|$OD)}}]{};

\draw[->,thick] (DO.north) .. controls +(up:8mm) and +(left:8mm) .. (OD.west) node[pos=0.4, label=above:{\rotatebox{45}{\parbox{1cm}{$. $\\[-1em]P(OD$|$DO)}}}]{};
\draw[->,thick] (OD.south west) .. controls +(left:8mm) and +(up:8mm) .. (DO.north east) node[pos=0.4, label=below:{\rotatebox{45}{\parbox{1cm}{$  $\\[-1em]P(DO$|$OD)}}}]{};

\draw[->,thick] (OD.south) .. controls +(down:0mm) and +(down:0mm) .. (DD.north) node[pos=0.3, label=below:{\rotatebox{-90}{\parbox{1cm}{$  $\\[2em]P(DD$|$OD)}}}]{};
\draw[->,thick] (DD.north east) .. controls +(down:0mm) and +(down:0mm) .. (OD.south east) node[pos=0.5, label=right:{\rotatebox{90}{P(OD$|$DD)}}]{};

\draw[->,thick] (DD.west) .. controls +(down:0mm) and +(down:0mm) .. (DO.east) node[pos=0.5, label=above:{\rotatebox{0}{P(DO$|$DD)}}]{};
\draw[->,thick] (DO.south east) .. controls +(down:0mm) and +(down:0mm) .. (DD.south west) node[pos=0.5, label=below:{\rotatebox{0}{P(DD$|$DO)}}]{};
\end{tikzpicture}
}

\caption{\label{fig:gnodeRep} Left: Push-pull gossip model of individual gossip node with rounds of length 3 (i.e. $\Gm=3$). Active states are red, passive ones blue. Model for replication. Right: Transition diagram of conditional probabilities pairwise interaction between gossip nodes. D: data element in cache; O: data element not in cache.}
\end{center}
\end{figure}

Without going into full detail\footnote{More details can be found in the Appendix, which will not be part of this paper.}, the mean field models proposed in the work by Bahkshi et al.~\cite{BakhshiCFH11} consider a gossip network as consisting of active and passive nodes that possess, or do not possess, the specific data element in their cache. This is illustrated in Fig.~\ref{fig:gnodeRep} (left), where the local states of a single node are shown. States in which the node actively looks for a gossip peer are red, those in which it passively receives requests are blue. States in which the node has the data-element in its cache are labelled by $D_i$, those in which it does not are labelled by $O_i$.  Transitions between states occur with certain probabilities, which depend on the global occupancy measure and the conditional probabilities of pairwise node interaction, shown in Fig.~\ref{fig:gnodeRep} (right), under the assumption of a uniform distribution of data items over the local storages of all nodes. $P(A'B'|AB)$ denotes the conditional probability of the state of an active-passive pair AB to have state $A'B'$ after their interaction, where $A,B,A',B' \in \{O,D\}$.

The conditional probabilities\footnote{See~\cite{Bak11,BakhshiCFH11} for further details on this pairwise communication probabilities.} can be expressed in terms of $n$ (number of different data elements), $c$ (size of the cache) and $s$ (number of selected elements for exchange), as follows:

\parbox{2in}{
$$
\begin{array}{l c l c l}
\mbox{P(OD$|$DO)} &= &\mbox{P(DO$|$OD)}& = & \frac{s}{c}*\frac{n-c}{n-s}\\
%\mbox{P(DO$|$OD)} & = &\mbox{P(OD$|$DO)}\\
\mbox{P(OD$|$OD)} &= &\mbox{P(DO$|$DO)} & = & \frac{c-s}{c}\\
%\mbox{P(DO$|$DO)} & = & \mbox{P(OD$|$OD)}\\
\mbox{P(DD$|$OD)} & = & \mbox{P(DD$|$DO)}  & = & \frac{s}{c}*\frac{c-s}{n-s}\\
\mbox{P(OD$|$DD)} & = & \mbox{P(DO$|$DD)} & = &\frac{s}{c}*\frac{c-s}{c}*\frac{n-c}{n-s}\\
\mbox{P(DD$|$DD)} &&& = & 1.0 -2.0*\frac{s}{c}*\frac{c-s}{c}*\frac{n-c}{n-s} \\
\mbox{P(OO$|$OO)}  &&& = & 1.0 
\end{array}
$$
}\quad
\raisebox{0.5em}{\parbox{2in}{
$$
\begin{array}{l c l c l}
%\mbox{P(OD$|$DD)} & = & \mbox{P(DO$|$DD)} & = &\frac{s}{c}*\frac{c-s}{c}*\frac{n-c}{n-s}\\
%\mbox{P(DD$|$DO)} & = & \mbox{P(DD$|$OD)}\\
%\mbox{P(DO$|$DD)} & = & \mbox{P(OD$|$DD)}\\
%\mbox{P(DD$|$DD)} & = & 1.0 -2.0*\frac{s}{c}*\frac{c-s}{c}*\frac{n-c}{n-s} &&\\
%\mbox{P(OO$|$OO)}  & = & 1.0 &&
\end{array}
$$
}
}

\begin{figure}[h] 
   \centering
  \includegraphics[width=2.0in]{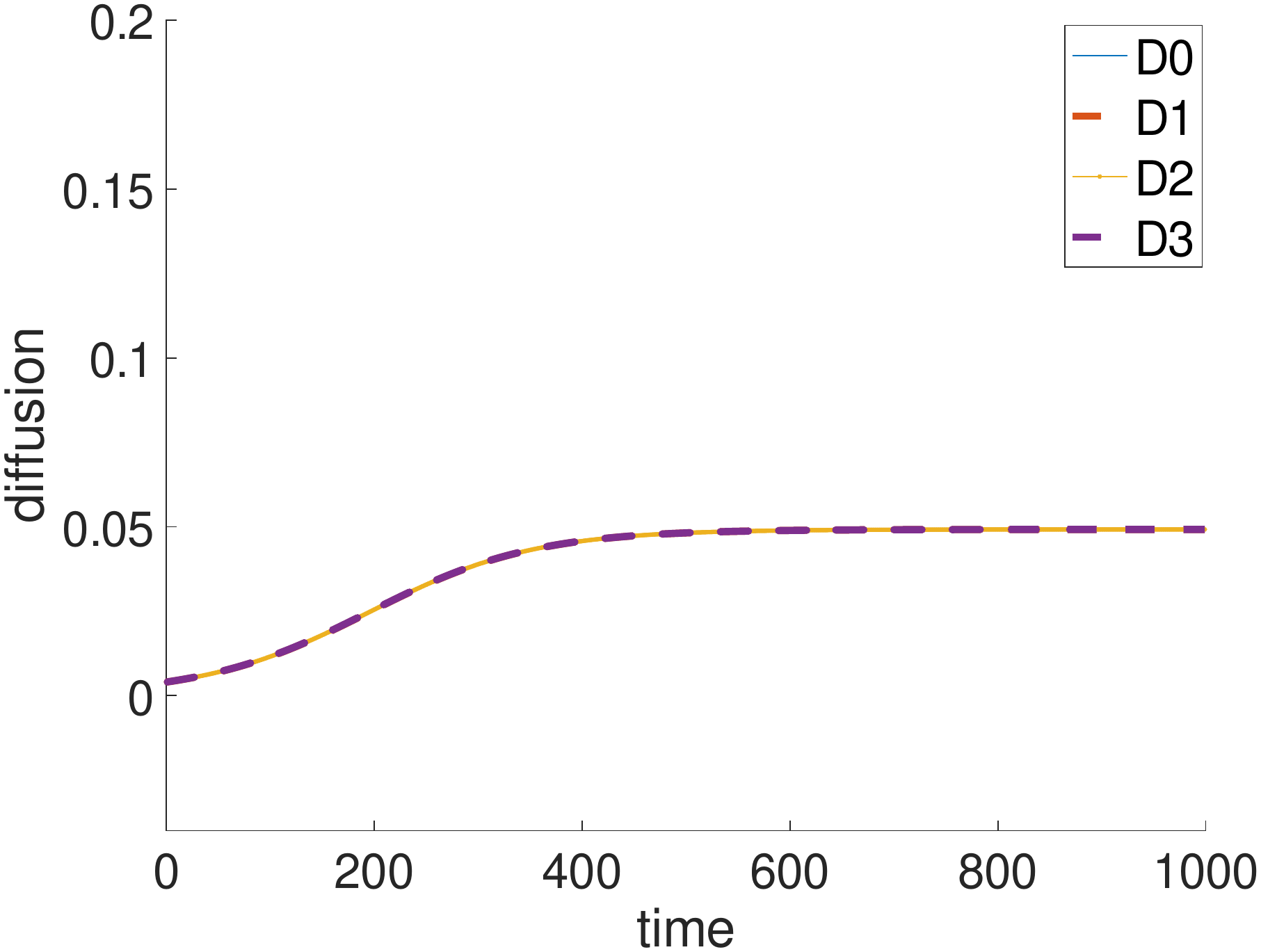}\quad\quad 
  \includegraphics[width=2.0in]{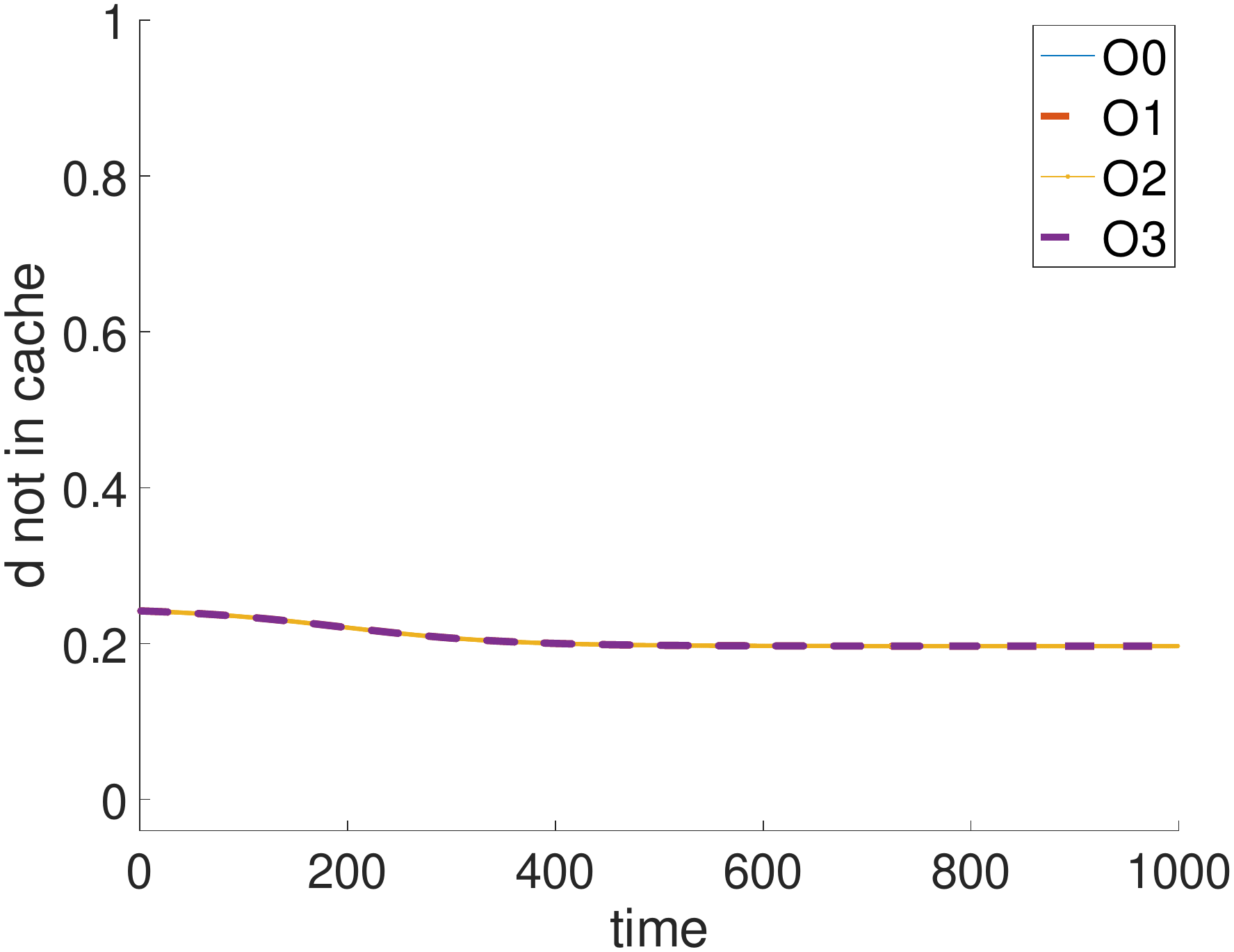} 
\caption{\label{fig:gossipN2500D10_equal_gm3}Diffusion of d-element in the network for $N=2500$ with initially 615 nodes in each O-state and 10 nodes in each D-state. Occupancy measure of D-states (left) and of O-states (right).}
\end{figure}

This mean field model can be further simplified, leading to a model that is parametric in $\Gm$, by aggregating the O-states and the D-states, respectively. This uses the experimental observation that when, in the initial state, the O-states all have the same occupancy measure, and all the D-states have the same occupancy measure, this situation remains so when time evolves\footnote{Note that we do {\em not} assume that the occupancy measure of an O-state is equal to that of a D-state.}. This observation is illustrated in Fig.~\ref{fig:gossipN2500D10_equal_gm3} for a network with 2500 nodes and $\Gm=3$, with 10 nodes in each O-state and 615 nodes in each D-state initially.

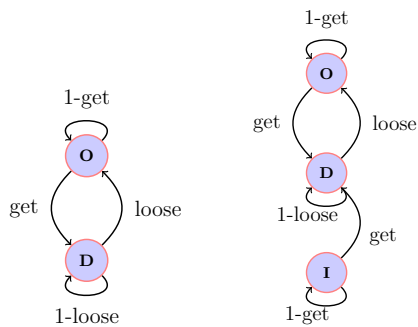
\begin{figure}
\begin{center}
\resizebox{0.21\textwidth}{!}{
\begin{tikzpicture}
\tikzstyle{place}=[circle,draw=blue!50,fill=blue!20,thick,inner sep=0pt,minimum size=8mm]
% Nodes with data-element
\node (D) [place,draw=red!50,fill=blue!20] at (0,0) {$\mathbf{D}$};

% Nodes without data-element
\node (O) [place,draw=red!50,fill=blue!20] at (0,2) {$\mathbf{O}$};s

%\draw[->,thick] (S.west) to [out=180,in=90] (S.north);
\draw[->,thick] (O.south west) ..controls +(-0.5,-0.5) and +(-0.5,0.5) ..(D.north west) node[pos=0.5, label=left:{{\large get}}]{};

\draw[->,thick] (D.north east) ..controls +(0.5,0.5) and +(0.5,-0.5) ..(O.south east) node[pos=0.5, label=right:{{\large loose}}]{};

\draw[->,thick] (O.north east) ..controls +(0.5,0.5) and +(-0.5,0.5) ..(O.north west) node[pos=0.5, label=above:{{\large 1-get}}]{};

\draw[->,thick] (D.south east) ..controls ++(0.5,-0.5) and +(-0.5,-0.5) ..(D.south west) node[pos=0.5, label=below:{{\large 1-loose}}]{};
\end{tikzpicture}
}\quad\quad
\resizebox{0.2\textwidth}{!}{
\begin{tikzpicture}
\tikzstyle{place}=[circle,draw=blue!50,fill=blue!20,thick,inner sep=0pt,minimum size=8mm]
% Nodes with data-element
\node (D) [place,draw=red!50,fill=blue!20] at (0,2) {$\mathbf{D}$};

% Nodes without data-element
\node (O) [place,draw=red!50,fill=blue!20] at (0,4) {$\mathbf{O}$};

% Nodes without data-element
\node (I) [place,draw=red!50,fill=blue!20] at (0,0) {$\mathbf{I}$};

%\draw[->,thick] (S.west) to [out=180,in=90] (S.north);
\draw[->,thick] (O.south west) ..controls +(-0.5,-0.5) and +(-0.5,0.5) ..(D.north west) node[pos=0.5, label=left:{{\large get}}]{};

\draw[->,thick] (D.north east) ..controls +(0.5,0.5) and +(0.5,-0.5) ..(O.south east) node[pos=0.5, label=right:{{\large loose}}]{};

\draw[->,thick] (O.north east) ..controls +(0.5,0.5) and +(-0.5,0.5) ..(O.north west) node[pos=0.5, label=above:{{\large 1-get}}]{};

\draw[->,thick] (D.south east) ..controls ++(0.5,-0.5) and +(-0.5,-0.5) ..(D.south west) node[pos=0.9, label=below:{{\large 1-loose}}]{};

\draw[->,thick] (I.north east) ..controls +(0.5,0.5) and +(0.5,-0.5) ..(D.south east) node[pos=0.3, label=right:{{\large get}}]{};

\draw[->,thick] (I.south east) ..controls ++(0.5,-0.5) and +(-0.5,-0.5) ..(I.south west) node[pos=0.9, label=below:{{\large 1-get}}]{};

%\draw[->,thick] (I.east) ..controls +(2,2) and +(0.7,-0.7) ..(O.east) node[pos=0.5, label=right:{{\large get\_exc}}]{};
%
\end{tikzpicture}
}
\caption{\label{fig:aggr_gnode_gm} Two-state (left) and three-state (right) aggregate push-pull gossip model of an individual gossip node with rounds of length $\Gm$.}
\end{center}
\end{figure}

 The simplified aggregated mean field model is shown in Fig.~\ref{fig:aggr_gnode_gm} (left) for the analysis of the replication, and on the right for the analysis of coverage. The latter shows an additional state (I). This models the state in which the node does not have the data-element in its cache currently and also has never had it before. The O-state in this model represents the fact that it does not have the data element at the moment, but that is has seen it previously (i.e. the node is already covered).
 
\label{page:TransProbFun}
The transition probability functions in the three-state model of Fig.~\ref{fig:aggr_gnode_gm}, with states $O$, $D$ and $I$ are defined as follows, for $m=(m_O,m_D) \in \calU^2$: % where $\calU^2$ is the unit simplex of $\nnreals^2$:
\begin{itemize}
\item from $O$ to $D$:\;%\\
%$\mbox{\sf get}$, with 
$\mbox{\sf get}\, (m) = \frac{\Gm}{\Gm+1}(\mbox{\sf ogs}'\, (m)) +\frac{1}{\Gm+1}(\mbox{\sf ogr}'\, (m))$%=
%$\frac{\Gm}{(\Gm+1)^2}\frac{s}{c}m_{Q_D}e^{-\frac{2}{\Gm+1}} +
%\frac{\Gm}{(\Gm+1)^2}\frac{s}{c}m_{Q_D}e^{-\frac{2}{\Gm+1}}$.
%$2\frac{\Gm}{(\Gm+1)^2}\frac{s}{c}m_{Q_D}e^{-\frac{2}{\Gm+1}}$.
%\item from $O$ to $O$:\;%\\
%$\mbox{\sf noget}$, with
%$\mbox{\sf noget}\, m = 1-(\mbox{\sf get}\, m)$.
\item from $D$ to $O$:\;%\\
%$\mbox{\sf loose}$, with
$\mbox{\sf loose}\, (m) =\frac{\Gm}{\Gm+1}(\mbox{\sf dls}'\, (m))+\frac{1}{\Gm+1}(\mbox{\sf dlr}'\, (m))$%=
%$\frac{\Gm}{\Gm+1}\frac{1}{\Gm+1}\frac{s}{c}\frac{n-c}{n-s}\frac{c-m_{Q_D}s}{c}e^{-\frac{2}{\Gm+1}}+
%\frac{1}{\Gm+1}\frac{\Gm}{\Gm+1}\frac{s}{c}\frac{n-c}{n-s}\frac{c-m_{Q_D}s}{c}e^{-\frac{2}{\Gm+1}}$.
%$2\frac{\Gm}{(\Gm+1)^2}\frac{s}{c}\frac{n-c}{n-s}\frac{c-m_{Q_D}s}{c}e^{-\frac{2}{\Gm+1}}$.
%\item from $D$ to $D$:\;%\\
%$\mbox{\sf keep}$, with
%$\mbox{\sf keep}\, m =1-(\mbox{\sf loose}\, m) $.
\item from $I$ to $D$:\;
$\mbox{\sf get}$ as above
\end{itemize}

\noindent
where\\
$\mbox{\sf ogs}' (m) =$
$\frac{1}{\Gm+1}m_{D}(P(OD|DO)+P(DD|DO))\mbox{\sf noc }$\\ %e^{-\frac{2}{\Gm+1}}$\\%=
%$\frac{1}{\Gm+1}\frac{s}{c}m_{D}e^{-\frac{2}{\Gm+1}}$;\\
$\mbox{\sf dls}' (m) =$
$\frac{1}{\Gm+1}((m_{O}+m_{I})P(DO|OD)+m_{D}P(DO|DD))\mbox{\sf noc }$\\ %e^{-\frac{2}{\Gm+1}}$\\%=
%$\frac{1}{\Gm+1}\frac{s}{c}\frac{n-c}{n-s}\frac{c-m_{D}s}{c}e^{-\frac{2}{\Gm+1}}$;\\
$\mbox{\sf ogr}' (m) =$
$\frac{\Gm}{\Gm+1}m_{D}(P(DO|OD)+P(DD|OD))\mbox{\sf noc }$\\ %e^{-\frac{2}{\Gm+1}}$\\%=
%$\frac{\Gm}{\Gm+1}\frac{s}{c}m_{D}e^{-\frac{2}{\Gm+1}}$; and\\
$\mbox{\sf dlr}'\, (m) =$
$\frac{\Gm}{\Gm+1}((m_{O}+m_{I})P(OD|DO)+m_{D}P(DO|DD))\mbox{\sf noc }$\\ %e^{-\frac{2}{\Gm+1}}$%=
%$\frac{\Gm}{\Gm+1}\frac{s}{c}\frac{n-c}{n-s}\frac{c-m_{D}s}{c}e^{-\frac{2}{\Gm+1}}$.

where $\mbox{\sf noc }$ is the no-collision probability, which, in the aggregated models, is equal to $e^{-2*(1/(\Gm+1))}$; note that $1/(\Gm+1)$ is the fraction of active nodes in the network at any time instant. This is derived from~\cite{Bak11}, where it is shown that in the limit for $N$ to infinity, the probability of no collision is given by $e^{-2*(\frc(O0)+\frc(D0))}$ where the sum $\frc(O0)+\frc(D0)$ denotes the fraction of active nodes in the network at any time. In the aggregated model this amounts to $1/(\Gm +1)$. In this model the number of replications of the data element in the network corresponds to the number of nodes that are in state D. The coverage of the network is given by the number of nodes that are in state D or state O.
%
% = e^{-2*(\frc(O0)+\frc(D0))}$ denotes the no-collision probability. 
%%In the limit for $N$ to infinity, the probability of no collision is given by $e^{-2*(\frc(O0)+\frc(D0))}$ where 
%The sum $\frc(O0)+\frc(D0)$ denotes the fraction of active nodes in the network at any time, i.e. in the aggregated model this amounts to $1/(\Gm +1)$. 
The definitions of the transition probabilities for the two-state model are similar, but with $m_{I}$ equal to zero. With the two state model only the number replications can be analysed. 

For very large systems both models show a surprisingly good correspondence between the Java simulation results and the classic mean field approximation. For N=25,000 the curves for both measures essentially overlap (see~\cite{Bak11,BakhshiCFH11}). For N=2,500, with initially one node in state D and all other nodes in state I,  for $\Gm=9$, the results for replication and coverage are shown in Fig.~\ref{fig:gossip3statesAgg}. For that system size already some differences can be observed, and, even though they are not huge, there is a considerable difference in the time at which network coverage seems to be reached. The Java simulation (average of 500 runs) indicates that this happened close to time 1500, whereas the mean field indicates a time well before that, just before time 1000, even though the mean field approximation is still just within the standard deviation of the simulation runs. In the next section we illustrate what results can be obtained with the refined mean field approximation and we also motivate why, in the general case, this requires a more detailed mean field model.

\section{Refined Mean Field Approximation of the Gossip Shuffle Protocol }
\label{sec:gossipRMF}

\begin{figure}[h] 
   \centering
   \includegraphics[width=2.0in]{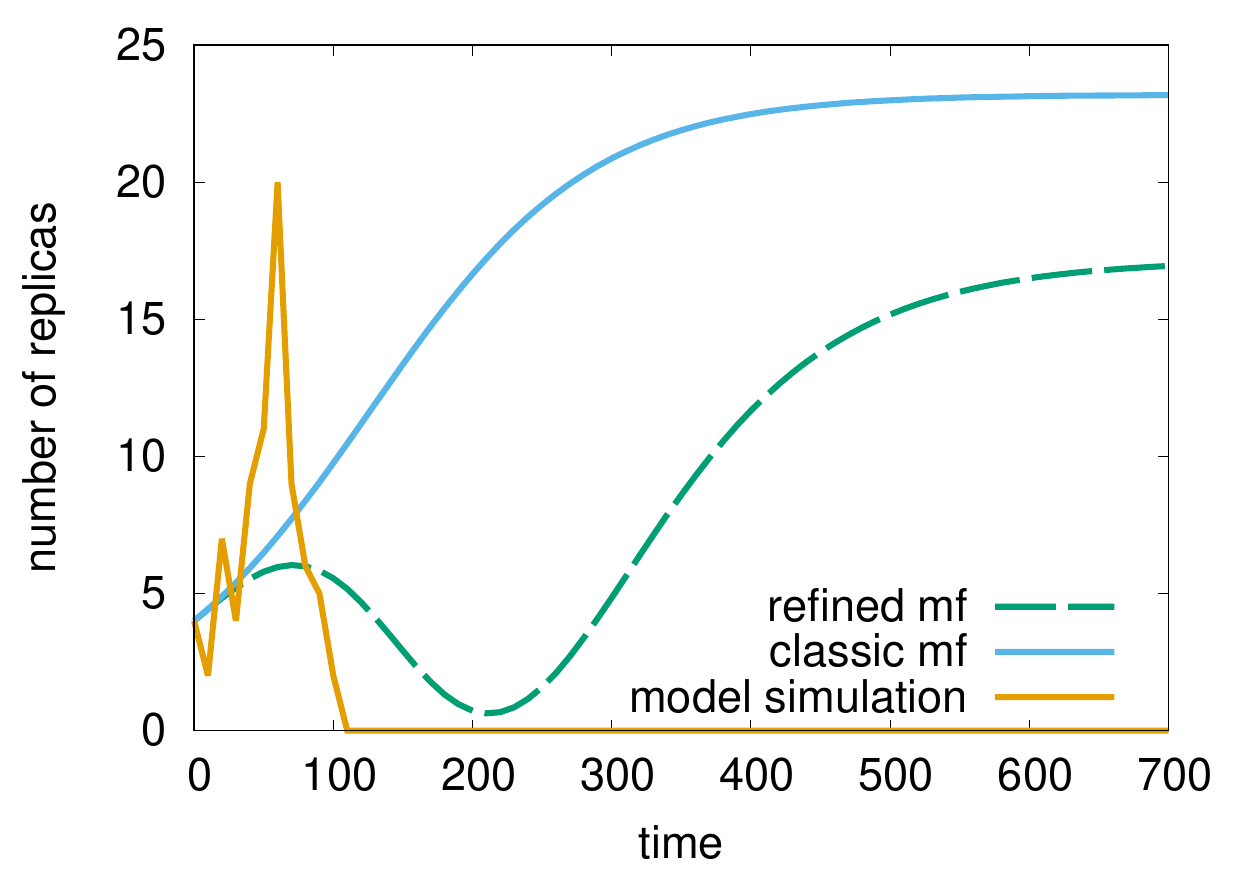}   
\caption{\label{fig:gossipN120D1sim1_full} Replication of data element in the network for $N=120$, $\Gm=3$, with initially 29 nodes in each O-state and 1 node in each D-state showing a single simulation trace of the model of Fig.~\ref{fig:gnodeRep} (left)in which the data element gets lost from the network. %, i.e. the number of replicas goes to zero just after 100 time units. 
The figure also shows the classic mean field (blue) and refined mean field (green) results.}
\end{figure}

The mean field models of the gossip shuffle protocol in the previous section were based on the principle of decoupling of joint probabilities~\cite{BMM07qest,BHLM13peva} based on a careful study of the pairwise probabilities of the various possible outcomes of a shuffle between two gossip nodes (as in~\cite{Bak11}). In our previous work on refined mean field approximation we have shown for a number of other models that this approximation technique can provide an increased accuracy w.r.t. classical mean field and that there is also a close correspondence between the simulation of the mean field model and the refined approximation~\cite{GLM18,GLM2019}. However, simulation of the {\em mean field model}\footnote{We really intend the simulation of the {\em model} here, and {\em not} the Java simulation of the protocol.} of Fig.~\ref{fig:gnodeRep} for a small network of size N=120, with $\Gm=3$, with initially  29 nodes in each O-state and one in each D-state, shows that in many simulation runs the system completely looses the introduced data-element. In other words, no gossip node in the network has the element in its cache at a certain point in time. This is clearly in contrast with the properties of the gossip protocol itself. The refined mean field approximation is also sensitive to this aspect of the model behaviour as can be observed in Fig.~\ref{fig:gossipN120D1sim1_full}. Similar observations can be made for the aggregated 3-state model of Fig.~\ref{fig:aggr_gnode_gm}. 

In the following we propose a more detailed mean field model in which (1) the system can never completely loose the inserted data element and (2) the model reflects more explicitly the {\em effects} of the pairwise interaction and synchronisation between nodes. Note the emphasis on {\em effects} of node synchronisation because we still are aiming at a model that respects the decoupling principle for its use in a mean field setting. What we really aim at is to distinguish the effects of a node getting a data element through {\em exchanging} it with another node--in which case the total number of replicas of the data element in the system remains the same--or through {\em replication}, i.e. the other node retains its copy of the data element and the global number of the data element in the system increases by one. 

\begin{figure}
\begin{center}
\resizebox{0.9\textwidth}{!}{
\begin{tikzpicture}
\tikzstyle{place}=[circle,draw=blue!50,fill=blue!20,thick,inner sep=0pt,minimum size=8mm]
% Objects with data-element
\node (D) [place,draw=red!50,fill=blue!20] at (0,2) {$\mathbf{D}$};

% Objects without data-element
\node (O) [place,draw=red!50,fill=blue!20] at (0,4) {$\mathbf{O}$};

% Objects with data-element for the first time
\node (FD) [place,draw=red!50,fill=blue!20] at (4,2) {$\mathbf{FD}$};

% Objects with data-element later-on
\node (LD) [place,draw=red!50,fill=blue!20] at (-4,4) {$\mathbf{LD}$};

% Object with permanent data-element
\node (PD) [place,draw=red!50,fill=blue!20] at (-4,0) {$\mathbf{PD}$};

% Initial objects without data-element
\node (I) [place,draw=red!50,fill=blue!20] at (0,0) {$\mathbf{I}$};

%\draw[->,thick] (S.west) to [out=180,in=90] (S.north);
\draw[->,thick] (O.south west) ..controls +(-0.5,-0.5) and +(-0.5,0.5) ..(D.north west) node[pos=0.5, label=left:{{\large get\_rep}}]{};

\draw[->,thick] (PD.south east) ..controls ++(0.5,-0.5) and +(-0.5,-0.5) ..(PD.south west) node[pos=0.5, label=below:{{\large 1}}]{};

\draw[->,thick] (D.north east) ..controls +(0.5,0.5) and +(0.5,-0.5) ..(O.south east) node[pos=0.5, label=right:{{\large loose\_rep}}]{};

\draw[->,thick] (O.north east) ..controls +(0.5,0.5) and +(-0.5,0.5) ..(O.north west) node[pos=0.5, label=above:{\parbox{2cm}{\large 1-(get\_rep)\\-(get\_exc)}}]{};

\draw[->,thick] (O.north west) ..controls +(-0.5,0.5) and +(0.5,0.5) ..(LD.north east) node[pos=0.5,label=above:{{\large get\_exc}}]{};

\draw[->,thick] (D.south east) ..controls ++(0.5,-0.5) and +(-0.5,-0.5) ..(D.south west) node[pos=0.9, label=below:{{\large 1-(loose\_rep)}}]{};

\draw[->,thick] (I.north east) ..controls +(0.5,0.5) and +(0.5,-0.5) ..(D.south east) node[pos=0.3, label=right:{{\large get\_rep}}]{};

\draw[->,thick] (LD.south) ..controls +(-0.0,-0.5) and +(-0.3,0.0) ..(D.west) node[pos=0.3, label=below:{{\large get\_rep}}]{};

\draw[->,thick] (LD.south east) ..controls +(-0.0,-0.5) and +(-0.5,-0.5) ..(O.south west) node[pos=0.5, label=above:{{\large loose\_exc}}]{};

\draw[->,thick] (I.east) ..controls +(0.5,0.0) and +(0.0,-0.8) ..(FD.south) node[pos=0.5, label=right:{{\large get\_exc}}]{};

\draw[->,thick] (FD.north) ..controls +(0.0,0.8) and +(0.5,-0.0) ..(O.east) node[pos=0.5, label=right:{{\large loose\_exc}}]{};

\draw[->,thick] (FD.west) ..controls +(0.0,0.0) and +(0.0,0.0) ..(D.east) node[pos=0.5, label=below:{{\large get\_rep}}]{};

\draw[->,thick] (FD.south east) ..controls ++(0.5,-0.5) and +(0.5,0.5) ..(FD.north east) node[pos=0.8, label=right:{\parbox{2.5cm}{\large 1-(loose\_exc)\\-(get\_rep)}}]{};

\draw[->,thick] (LD.north west) ..controls +(-0.5,0.5) and +(-0.5,-0.5) ..(LD.south west) node[pos=0.5, label=left:{\parbox{2.5cm}{\large 1-(loose\_exc)\\ - (get\_rep)}}]{};

\draw[->,thick] (I.south east) ..controls ++(0.5,-0.5) and +(-0.5,-0.5) ..(I.south west) node[pos=0.9, label=below:{\parbox{2cm}{\large 1-(get\_exc)\\-(get\_rep)}}]{};

%\draw[->,thick] (I.east) ..controls +(2,2) and +(0.7,-0.7) ..(O.east) node[pos=0.5, label=right:{{\large get\_exc}}]{};
%
\end{tikzpicture}
}
\caption{\label{fig:aggr_gnode_noloss_sync_6states} Six-state model of an individual gossip node with rounds of length $\Gm$.
}
\end{center}
\end{figure}
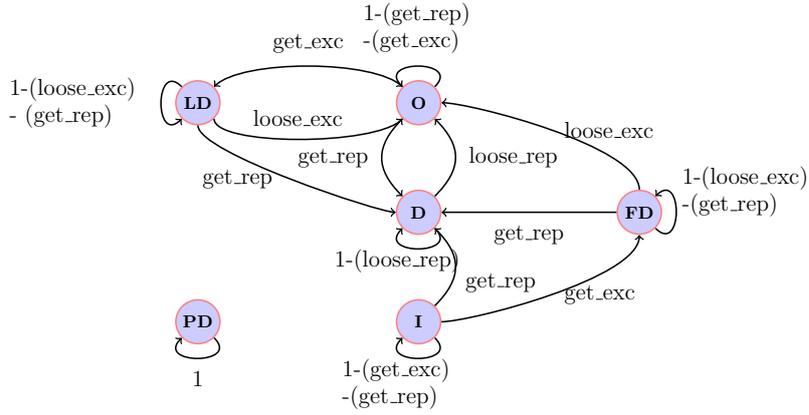

With reference to Fig.~\ref{fig:aggr_gnode_noloss_sync_6states}, for what concerns point (1) above, we introduce a specific state, {\sf PD}, to the model representing that there always is a gossip node in the network that possesses the data element. 
%Note that this does not imply that this node is any specific node since the mean field model abstracts from the identity of objects. Hence, such a node with the data element only reflects the fact that there is always at least one node in the network that has the data element in its cache.

To address point (2), we introduce two more states, {\sf FD} and {\sf LD}, to distinguish between the effect of interactions between gossip nodes. State {\sf FD} represents the fact that the gossip node received the data element for the {\em first time} via an {\em exchange} of the data element with another node. State {\sf LD} also represents the fact that the node received the data element via an {\em exchange}, but that it had already seen the data element in the past. So in both cases, the data element is simply exchanged, i.e. one node gives it to the other, and the total number of gossip nodes that possess the data element is not changed by such an interaction. Note that modelling the effect of an exchange of the data element between two nodes in this way also means that we can retrieve the total number of gossip nodes in the system that do {\em not} possess the data element as {\em the sum} of the nodes that are in states FD, LD, I and O. This is so because we know that for each node in state FD (LD, resp.) there is a node in the network that just lost its data element in the synchronous shuffle with our current node. We will make use of this in the probability functions associated with the transitions between nodes.

A gossip node can also get involved in an interaction in which the data element is replicated, i.e. a node gives it to another one but also retains a copy itself. Note that this can happen both in case the node that receives the data element does not possess the data element and when it does possess it. This situation is modelled by state {\sf D} and represents the fact that the interaction has the effect that the total number of nodes in the network that possess the data element increases (by one).

A third case exists where two nodes, both possessing the data element, interact and one of them looses its copy. In that case the overall number of copies of the data element in the network is reduced by one. Note that the gossip protocol does not allow that {\em both} copies get lost in such an interaction. Moreover, if there is only a single node left in the network with a copy of the data element this copy cannot get lost because this node cannot interact with another node having the data element.

To distinguish the various kinds of interactions mentioned above we refine the transition probability functions introduced on page~\pageref{page:TransProbFun}. In particular, we split the probability functions $\mbox{\sf get}$ and $\mbox{\sf loose}$ into two distinct parts, $\mbox{\sf get\_rep}$ and $\mbox{\sf get\_exc}$ for the $\mbox{\sf get}$ function to model data element replication and exchange, respectively, and likewise for the $\mbox{\sf loose}$ function as follows, where the appropriate conditional probabilities are used:
$$
\begin{array}{l c l}
\mbox{\sf get\_exc}\, (m)  &=& 2*\frac{\Gm}{(\Gm+1)^2}(m_{D}+m_{\mathit{PD}})P(OD|DO)\mbox{\sf noc }\\ %e^{-\frac{2}{\Gm+1}}\\
\mbox{\sf get\_rep}\, (m)  &=& 2*\frac{\Gm}{(\Gm+1)^2}(m_{D}+m_{\mathit{PD}})P(DD|DO)\mbox{\sf noc }\\[1em] %e^{-\frac{2}{\Gm+1}}\\[1em]
\mbox{\sf loose\_exc}\, (m)  &=& 2*\frac{\Gm}{(\Gm+1)^2}(m_{O}+m_{I}+m_{\mathit{LD}}+m_{\mathit{FD}})P(OD|DO)\mbox{\sf noc }\\ %e^{-\frac{2}{\Gm+1}}\\
\mbox{\sf loose\_rep}\, (m)  &=& 2*\frac{\Gm}{(\Gm+1)^2}(m_{D}+m_{\mathit{PD}})P(DO|DD)\mbox{\sf noc }\\ %e^{-\frac{2}{\Gm+1}}\\
\end{array}
$$
%

%In summary, in the model presented in Fig.~\ref{fig:aggr_gnode_noloss_sync_6states}, state {\sf I} models that the node has not yet received the data element. It can obtain the data element in two ways: by exchange with a node that has it or by replication of the element when interacting with a node that has it. In the first case it goes to state {\sf FD} with a transition labelled by {\sf get\_exc}, denoting the probability to get the data element by exchange. In the second case it goes to state {\sf D} with a transition labelled by {\sf get\_rep}, denoting the probability of replication of the data-element. If the node is in state {\sf FD} it has the data element and can loose it in an exchange with another node with probability {\sf loose\_exc} going to state {\sf O} denoting that it no longer has the data element, but that it has seen it in the past. From {\sf FD} it can also go to {\sf D} denoting that it interacted with another node and replicated the data element, increasing the total number of nodes having the data element. From {\sf O} it can again obtain the data element by exchange {\sf get\_exc}, or by replication {\sf get\_rep}, going respectively to state {\sf LD} or to state {\sf D}. State {\sf LD} denotes that the node has the data element, that it was obtained by exchange, but that it is not the first time it receives the data element. Once in {\sf LD} it can again loose the element in an exchange, returning to state {\sf O}, or get involved in an interaction with a node that does not possess the data element and replicate it, going to state {\sf D}.

\begin{figure} [t!!!]
%\centerline{
 \includegraphics[width=0.49\textwidth]{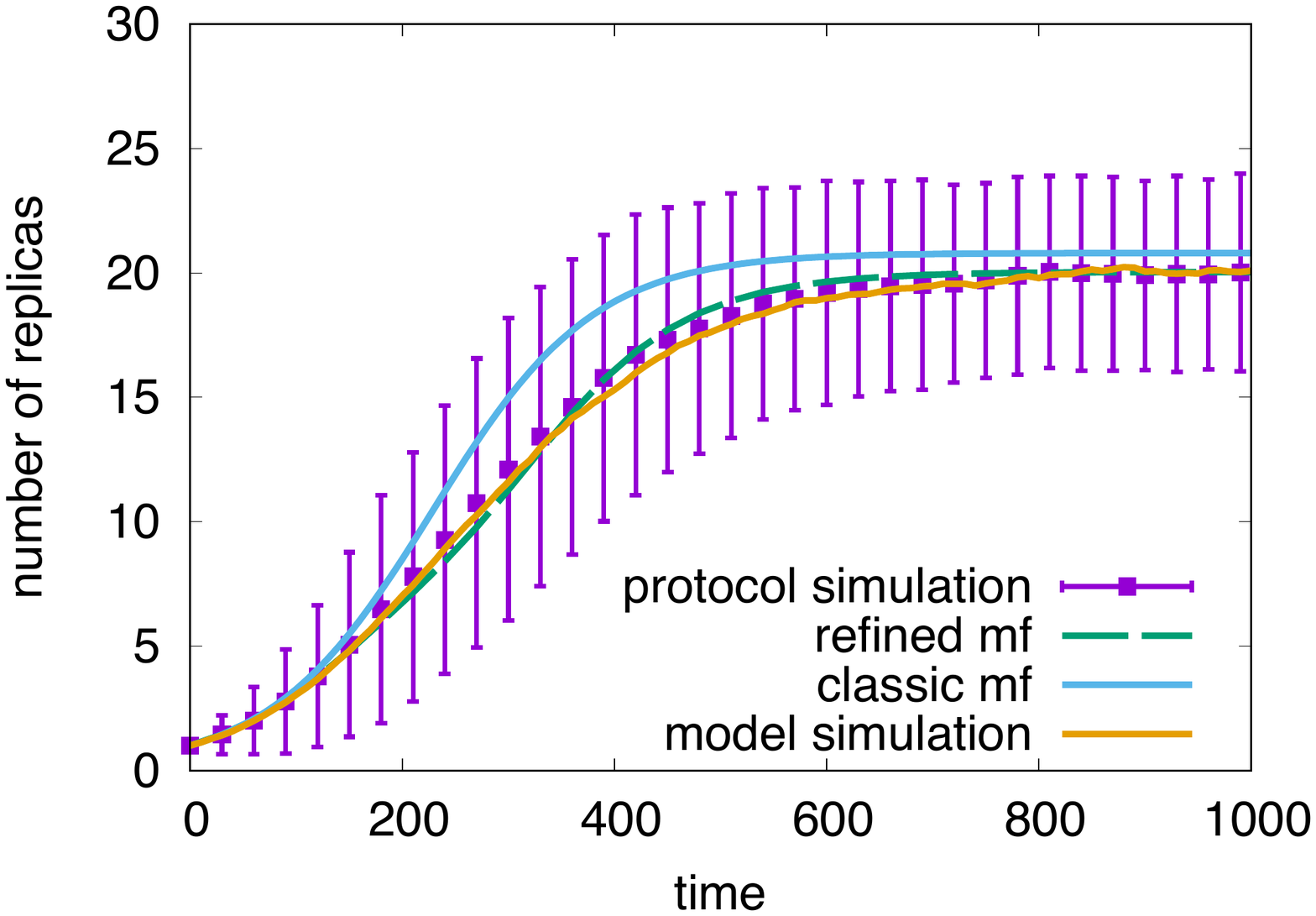}
 \includegraphics[width=0.49\textwidth]{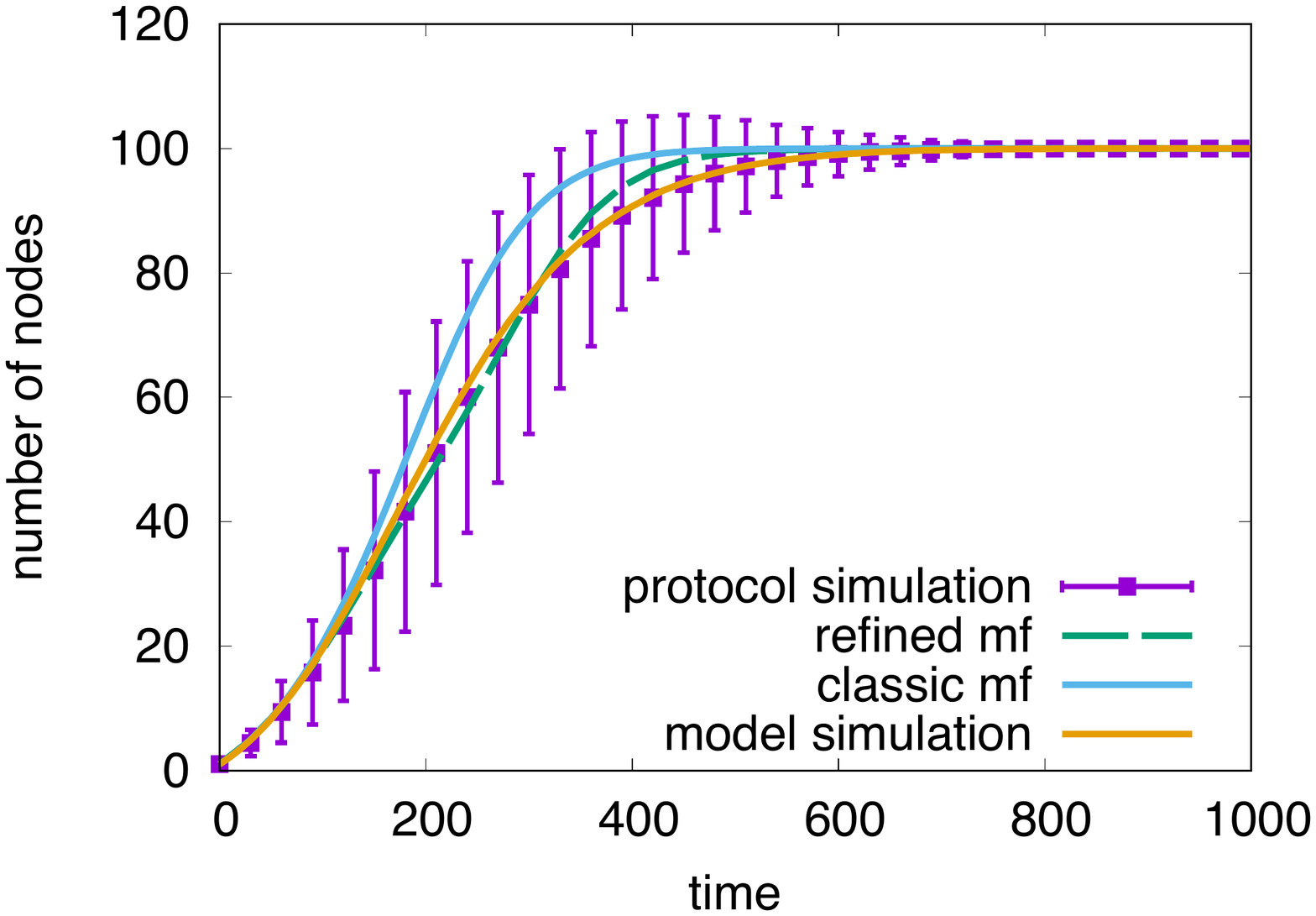}%\\[-7em]
%}
\caption{\label{fig:gossip5+1statesyncN23D1gmax9small}Replication (left) and network coverage (right) of the data element in the network for $N=100$ with initially 99 nodes in the {\sf I}-state and 1 node in the {\sf PD}-state for $\Gm=3$. Average of 500 simulation runs of both the model and Java simulations. Vertical bars show standard deviation for the Java simulation.}
\end{figure}

\begin{figure}[h!!!]
%   \centerline{
    \includegraphics[width=0.49\textwidth]{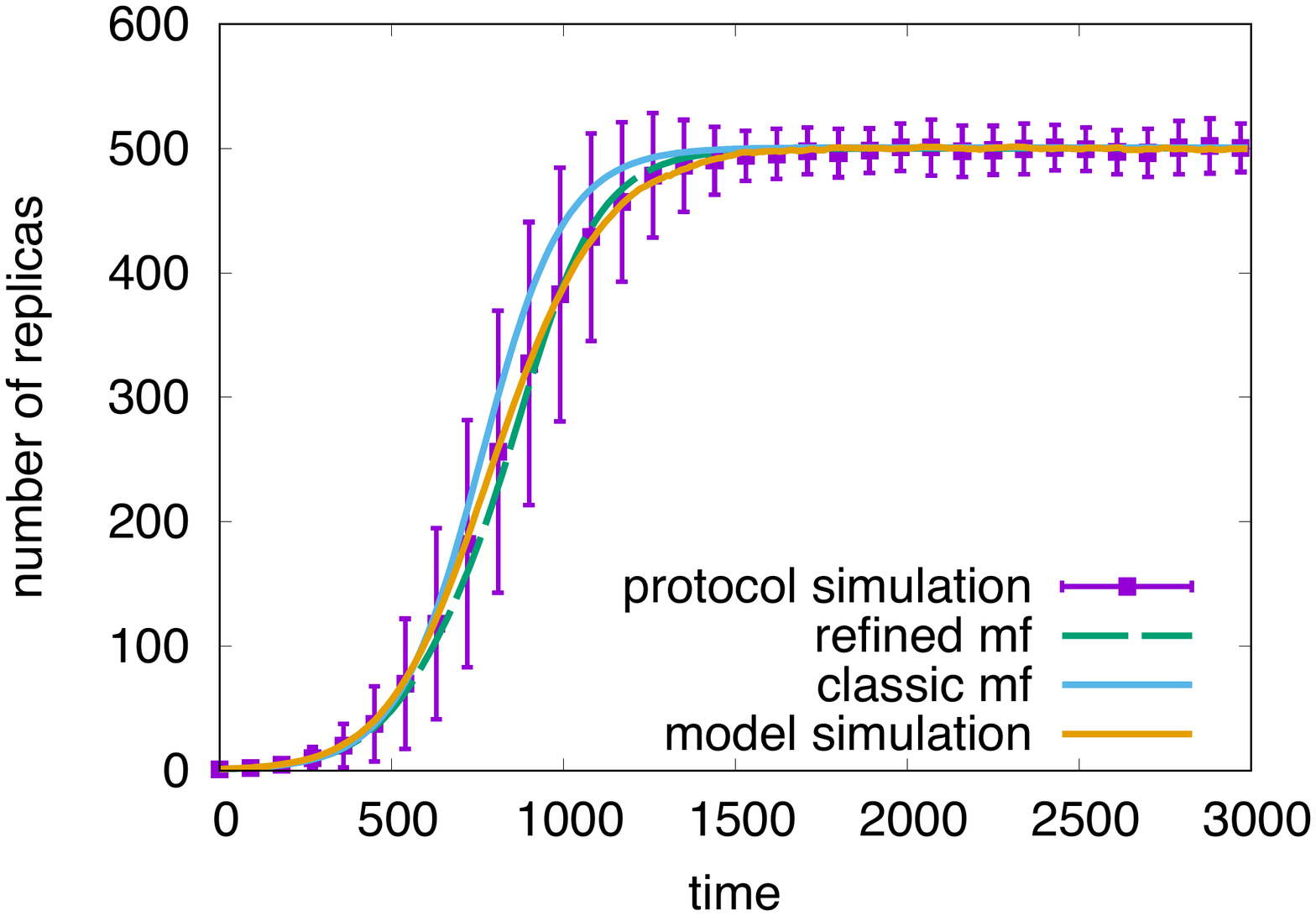}
    \includegraphics[width=0.49\textwidth]{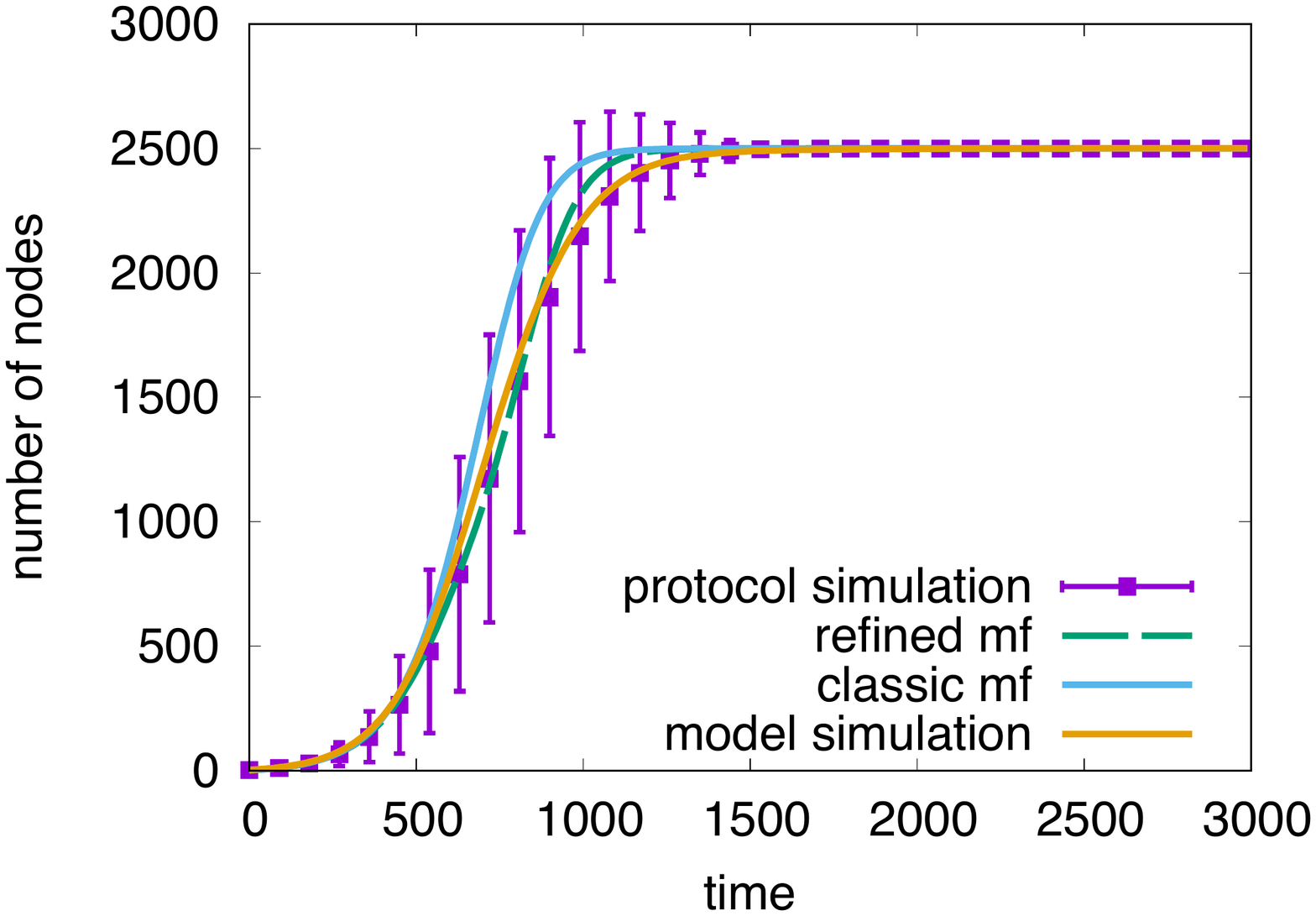}\\[-7em]
    %}
\caption{\label{fig:gossip3statesyncN2500D1gmax9}Replication (left) and network coverage (right) of data element for $N=2500$ with initially 2499 nodes in  {\sf I} and 1  in  {\sf PD}, for $\Gm=9$. Average of 500 simulation runs for both model and Java simulations. Vertical bars show standard deviation for the Java simulation.}
\end{figure}

Fig.~\ref{fig:gossip5+1statesyncN23D1gmax9small} shows the replication as sum of the number of nodes in states D and PD and the coverage as the sum of the number of nodes in D, PD, FD, LD and O\footnote{For the refined mean field this means the application of Thm. 1 with $h(m)= m_D+m_{PD}$ (replication) and $h(m)=m_D+m_{PD}+m_{FD}+m_{LD}+m_{O}$ (coverage), repectively.} for a network with $N=100$, $n=500$, $c=100$ and $s=50$ with initially one node in state PD and all the others in state I. Besides the classic and refined mean field approximations for the model in Fig.~\ref{fig:aggr_gnode_noloss_sync_6states} and the Java simulation results of the actual shuffle protocol, Fig.~\ref{fig:gossip5+1statesyncN23D1gmax9small} also shows the average of the model simulation. In particular, note the good approximation of the simulation results (both the Java simulation and the model simulation) by the refined mean field even in this very small network. This holds both for the diffusion of the replicas and for the coverage. 
%This confirms our conjecture that a model that better captures the effect of coupled transitions of a pair of nodes when they shuffle considerably improves the approximation results.
%
%
Similarly good results have been found for a system with N=2,500 shown in Fig.~\ref{fig:gossip3statesyncN2500D1gmax9}, also in the case in which there is only a single data element in the system initially.
An indication of the (non-optimised) performance of the analysis for producing the results in Fig.~\ref{fig:gossip3statesyncN2500D1gmax9} is: 0.543s (classic mean field); 25.459s (refined mean field); 7m 1.389s (fast model simulation~\cite{BMM07qest}, 500 runs); 3h 42m 41.459s (Java simulation, 500 runs) on a MacBook Pro, Intel i7, 16GB. Recall that the mean field and refined mean field analyses times are {\em independent} of the size of the system and, as can be seen, several orders of magnitude faster than traditional event simulation approaches. % even when starting from a single node that initially has the data element.

\section{Conclusion}
\label{sec:concl}

Gossip protocols play an important role in the design of collective adaptive systems providing a basic, but robust and scalable, mechanism of information spreading in very large networks. Therefore they also form an interesting benchmark application for the analysis of scalable verification techniques. We have developed a new mean field model for the shuffle gossip protocol with which more accurate approximations for medium size gossip protocols can be obtained via refined mean field approximation techniques. This model respects key aspects of the protocol such as the effects of different kinds of interactions and the fact that a new data element cannot be lost by the system as a whole. 

Good approximation of medium size systems is of interest for several reasons. First of all, many practical systems consist of many, but not a huge number, of components. However, even in case of medium size systems, simulation is still a resource consuming effort and in that case a refined mean field approximation can provide fast but accurate approximations. Furthermore, we expect that refined mean field approximation can also be of use when analysing systems in which objects are mobile and move through physical space. The uneven distribution of objects over partitions of such a space requires a mean field approximation that is accurate also for those partitions with relatively few objects. 

\section{Acknowledgements} We wish to thank Rena Bakhshi for sharing with us her Java simulator software for the reproduction of the gossip protocol simulations.

This research has been partially supported by the MIUR project PRIN 2017FTXR7S ``IT-MaTTerS" (Methods and Tools for Trustworthy Smart Systems).

\bibliographystyle{splncs03}
\bibliography{coordination19.bib,RefMeanField.bib}
% !TEX root =  QAPL17_FF.tex

%\newpage
\section{Appendix: Detailed Models and Proofs}

\subsection{Details for gossip model in Fig.~\ref{fig:gnodeRep}}

Fig.~\ref{fig:gnodeRep} (left) shows the states and transitions of a single gossip node where $\Gm=3$. The red states, $D0$ and $O0$, denote states in which the gossip node is {\em active}, i.e. it can initiate an exchange of local information with a passive node; in $D0$ (resp. $O0$) the node has (resp. does not have) the data element in its local cache. The blue states denote states in which the node is passive and is available for data exchange with an active node when contacted by the latter. The number in the node-labels denotes the value, ranging from 0 to 3, of the current gossip delay $g$ before the node becomes active again. The $D/O$ convention w.r.t. having the data element applies also to the states where the node is passive. 
The transition labels in Fig.~\ref{fig:gnodeRep} (left) are shorthands for transition probability functions. The latter depend on the occupancy measure vector. Their definition makes use of the conditional probabilities shown in Fig.~\ref{fig:gnodeRep} (right). Furthermore, we recall from~\cite{Bak11} that there is a small probability that collision occurs in the communication between two nodes. This happens when a gossip partner is selected that is already involved in a shuffle with another node. In the limit for $N$ to infinity, the value of the probability of no collision is given by $e^{-2*(\frc(O0)+\frc(D0))}$ where $\frc(O0)+\frc(D0)$ denotes the fraction of active nodes in the network at any time, i.e. summing active nodes that have the d-element and those that do not.

In the definition of the transition probability functions, we also make use of the following observation that greatly simplifies the definitions. Note that this gossip model is a clock-synchronous model and that each node gets active every $\Gm$ time steps. This means that in every step, if the node is in state $Di$ (or $Oi$) then in the next step it leaves this state with probability 1.0 to move one step closer towards the active state $D0$, or, if it was active, it moves to $D3$ (or $O3$) modelling a reset of the time-to-activation; similarly for $Oi$ states. In other words, for all time steps $t$ it holds that:
$$
\mathsf{dks}(\mu(t)) + \mathsf{dls}(\mu(t)) =1 \mbox{ and }\\
\mathsf{ons}(\mu(t)) + \mathsf{ogs}(\mu(t)) =1
$$
Similarly for the reset probabilities. The proofs can be found in Sect.~\ref{sec:proofs} of this Appendix. 

%For ease of notation we introduce occupancy measure vector $m$ such that $m_0=\frc(O0)$ to $m_3=\frc(O3)$ denote the fraction of nodes in state $O0$ to $O3$, respectively, and $m_4=\frc(D0)$ to $m_7=\frc(D3)$ denote the fraction of nodes in state $D0$ to $D3$, respectively. 
The transition probability functions that concern the $D$-states are defined as shown below. Note that here and in the remainder of the Appendix we also use Currying notation for notational simplicity. In the sequel $m$ denotes the occupancy measure {\em vector}. Its components are indicated by $m_{O1}, m_{O2}$ and so on.

$$
\begin{array}{l c l}
\mbox{\sf d\_loss\_reset}\;  m& = & (m_{O1}+m_{O2}+m_{O3})*P(OD|DO)*(\mbox{\sf noc })+\\ % e^{-2*(m_0+m_4)}+\\
 & &                                             (m_{D1}+m_{D2}+m_{D3})*P(OD|DD)*(\mbox{\sf noc })\\[1em] %e^{-2*(m_0+m_4)}\\[1em]
\mbox{\sf d\_loss\_step}\;  m& = & m_{O0}*P(DO|OD)*(\mbox{\sf noc })+%\\ %e^{-2*(m_0+m_4)}+\\
%& &                                             
m_{D0}*P(DO|DD)*(\mbox{\sf noc })\\[1em] %e^{-2*(m_0+m_4)}\\[1em]
\mbox{\sf d\_keep\_reset}\;  m& = & 1.0 - (\mbox{\sf d\_loss\_reset}\;  m)\\[1em]
\mbox{\sf d\_keep\_step}\;  m& = & 1.0 - (\mbox{\sf d\_loss\_step}\;  m)\\[1em]
\end{array}
$$
Function $\mbox{d\_loss\_reset}$ corresponds to the transition $\mbox{dlr}$ in Fig.~\ref{fig:gnodeRep}, and so on.

The transition probability functions concerning the $O$-states are:

$$
\begin{array}{l c l}
\mbox{\sf o\_getd\_reset}\; m & = & (m_{D1}+m_{D2}+m_{D3})*(P(DO|OD) + P(DD|OD))*(\mbox{\sf noc })\\[1em] %e^{-2*(m_0+m_4)}\\[1em]
\mbox{\sf o\_getd\_step}\; m & = & m_{D0}*(P(OD|DO) + P(DD|DO))*(\mbox{\sf noc })\\[1em] %e^{-2*(m_0+m_4)}\\[1em]
\mbox{\sf o\_nod\_reset}\; m & = & 1.0-(\mbox{\sf o\_getd\_reset}\; m)\\[1em]
\mbox{\sf o\_nod\_step}\; m & = & 1.0 - (\mbox{\sf o\_getd\_step}\; m)
\end{array}
$$

Defining a 8 $\times$ 8 matrix $\vr{K}$ with indexes $i,j$ in $\{0,\cdots, 7\}$ then we can define the gossip model as follows for the non-zero elements of $\vr{K}$:

$$
\begin{array}{l c l }
K_{0,3} & = & 1.0-(\mbox{\sf o\_getd\_reset}\; m)\\
K_{0,7} & = & \mbox{\sf o\_getd\_reset}\; m\\
K_{1,0}  =  K_{2,1}  =  K_{3,2} & = & 1.0 - (\mbox{\sf o\_getd\_step}\; m)\\
K_{1,4}  =  K_{2,5} = K_{3,6} & = & \mbox{\sf o\_getd\_step}\; m\\[1em]
K_{4,3} &=& \mbox{\sf d\_loss\_reset}\;  m\\
K_{4,7} &=& 1.0 - \mbox{\sf d\_loss\_reset}\;  m\\
K_{5,0} = K_{6,1} = K_{7,2} & = & \mbox{\sf d\_loss\_step}\;  m\\
K_{5,4} = K_{6,5} = K_{7,6} & = & 1.0 - \mbox{\sf d\_loss\_step}\;  m
\end{array}
$$
This leads a set of eight difference equations for the model recalling that the occupancy vector $\vt{m}$ at time $t+1$ is $\vt{m}(t+1) = \vt{m}(t)*\vr{K}(\vt{m}(t))$. These equations are shown in full in the next section.

\subsection{Difference equations for the gossip model in Fig.~\ref{fig:gnodeRep}} % Fig.~\ref{fig:gnode}}

We obtain the following set of difference equations\footnote{\label{note} On the left of each equation the new value of the occupancy measure vector at time step $t+1$, on the right the values of $m$ are intended to be those at time $t$. For notational simplicity $t+1$ and $t$ have been omitted in the equations below.} for the O-states and the D-states of the model, representing the occupancy measure of state Oi by $m_{Oi}$, and Di by $m_{Di}$, for $i \in \{0,\cdots,3\}$.
%O1 by $m_{O1}$, O2 by $m_{O2}$, O3 by $m_{O3}$, 
%D0 by $m_{D0}$, D1 by $m_{D1}$, D2 by $m_6$ and D3 by $m_7$.

$$
\begin{array}{l c l}
m_{O0} & = &  m_{O1} - m_{O1}*(\mbox{\sf o\_getd\_step}\; m) +m_{D1}*(\mbox{\sf d\_loss\_step}\; m)\\
m_{O1} & = &  m_{O2} - m_{O2}*(\mbox{\sf o\_getd\_step}\; m) +m_{D2}*(\mbox{\sf d\_loss\_step}\; m)\\
m_{O2} & = &  m_{O3} - m_{O3}*(\mbox{\sf o\_getd\_step}\; m) +m_{D3}*(\mbox{\sf d\_loss\_step}\; m)\\
m_{O3} & = &  m_{O0} - m_{O0}*(\mbox{\sf o\_getd\_reset}\; m) +m_{D0}*(\mbox{\sf d\_loss\_reset}\; m)\\[1em]
m_{D0} & = &  m_{O1}*(\mbox{\sf o\_getd\_step}\; m) +m_{D1}- m_{D1}*(\mbox{\sf d\_loss\_step}\; m)\\
m_{D1} & = &  m_{O2}*(\mbox{\sf o\_getd\_step}\; m) +m_{D2} - m_{D2}*(\mbox{\sf d\_loss\_step}\; m)\\
m_{D2} & = &  m_{O3}*(\mbox{\sf o\_getd\_step}\; m) +m_{D3} - m_{D3}*(\mbox{\sf d\_loss\_step}\; m)\\
m_{D3} & = &  m_{O0}*(\mbox{\sf o\_getd\_reset}\; m) +m_{D0} - m_{D0}*(\mbox{\sf d\_loss\_reset}\; m)
\end{array}
$$

\subsection{Classic Mean Field Model: Coverage.}

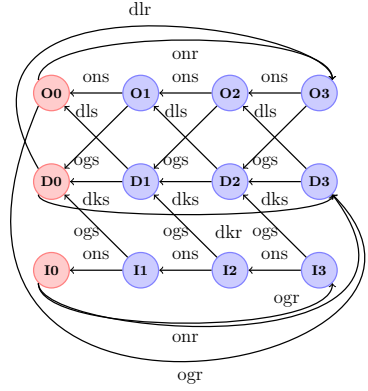
\begin{figure}[h!]
\begin{center}
\resizebox{0.7\textwidth}{!}{
\begin{tikzpicture}
\tikzstyle{place}=[circle,draw=blue!50,fill=blue!20,thick,inner sep=0pt,minimum size=8mm]
% Nodes with data-element
\node (D0) [place,draw=red!50,fill=red!20] at (0,2) {$\mathbf{D0}$};
\node (D1) [place,draw=blue!50,fill=blue!20] at (2,2) {$\mathbf{D1}$};
\node (D2) [place,draw=blue!50,fill=blue!20] at (4,2) {$\mathbf{D2}$};
\node (D3) [place,draw=blue!50,fill=blue!20] at (6,2) {$\mathbf{D3}$};

% Nodes without data-element, but that have seen data element before
\node (O0) [place,draw=red!50,fill=red!20] at (0,4) {$\mathbf{O0}$};
\node (O1) [place,draw=blue!50,fill=blue!20] at (2,4) {$\mathbf{O1}$};
\node (O2) [place,draw=blue!50,fill=blue!20] at (4,4) {$\mathbf{O2}$};
\node (O3) [place,draw=blue!50,fill=blue!20] at (6,4) {$\mathbf{O3}$};

% Nodes without data-element and have never seen it before
\node (IO0) [place,draw=red!50,fill=red!20] at (0,0) {$\mathbf{I0}$};
\node (IO1) [place,draw=blue!50,fill=blue!20] at (2,0) {$\mathbf{I1}$};
\node (IO2) [place,draw=blue!50,fill=blue!20] at (4,0) {$\mathbf{I2}$};
\node (IO3) [place,draw=blue!50,fill=blue!20] at (6,0) {$\mathbf{I3}$};

%\draw[->,thick] (S.west) to [out=180,in=90] (S.north);
\draw[->,thick] (D3.west) .. controls +(up:0mm) and +(up:0mm) .. (D2.east) node[pos=0.5, label=below:{{\large dks}}]{};
\draw[->,thick] (D2.west) .. controls +(up:0mm) and +(up:0mm) .. (D1.east) node[pos=0.5, label=below:{{\large dks}}]{};
\draw[->,thick] (D1.west) .. controls +(up:0mm) and +(up:0mm) .. (D0.east) node[pos=0.5, label=below:{{\large dks}}]{};

\draw[->,thick] (O3.west) .. controls +(up:0mm) and +(up:0mm) .. (O2.east) node[pos=0.5, label={{\large ons}}]{};
\draw[->,thick] (O2.west) .. controls +(up:0mm) and +(up:0mm) .. (O1.east) node[pos=0.5, label={{\large ons}}]{};
\draw[->,thick] (O1.west) .. controls +(up:0mm) and +(up:0mm) .. (O0.east) node[pos=0.5, label={{\large ons}}]{};

\draw[->,thick] (IO3.west) .. controls +(up:0mm) and +(up:0mm) .. (IO2.east) node[pos=0.5, label={{\large ons}}]{};
\draw[->,thick] (IO2.west) .. controls +(up:0mm) and +(up:0mm) .. (IO1.east) node[pos=0.5, label={{\large ons}}]{};
\draw[->,thick] (IO1.west) .. controls +(up:0mm) and +(up:0mm) .. (IO0.east) node[pos=0.5, label={{\large ons}}]{};

\draw[->,thick] (D3.north west) .. controls +(up:0mm) and +(up:0mm) .. (O2.south east) node[pos=0.6, label={{\large dls}}]{};
\draw[->,thick] (D2.north west) .. controls +(up:0mm) and +(up:0mm) .. (O1.south east) node[pos=0.6, label={{\large dls}}]{};
\draw[->,thick] (D1.north west) .. controls +(up:0mm) and +(up:0mm) .. (O0.south east) node[pos=0.6, label={{\large dls}}]{};

\draw[->,thick] (O3.south west) .. controls +(up:0mm) and +(up:0mm) .. (D2.north east) node[pos=0.6, label=below:{{\large ogs}}]{};
\draw[->,thick] (O2.south west) .. controls +(up:0mm) and +(up:0mm) .. (D1.north east) node[pos=0.6, label=below:{{\large ogs}}]{};
\draw[->,thick] (O1.south west) .. controls +(up:0mm) and +(up:0mm) .. (D0.north east) node[pos=0.6, label=below:{{\large ogs}}]{};

\draw[->,thick] (IO3.north west) .. controls +(up:0mm) and +(up:0mm) .. (D2.south east) node[pos=0.6, label=below:{{\large ogs}}]{};
\draw[->,thick] (IO2.north west) .. controls +(up:0mm) and +(up:0mm) .. (D1.south east) node[pos=0.6, label=below:{{\large ogs}}]{};
\draw[->,thick] (IO1.north west) .. controls +(up:0mm) and +(up:0mm) .. (D0.south east) node[pos=0.6, label=below:{{\large ogs}}]{};

\draw[->,thick] (O0.north west) .. controls +(up:12mm) and +(up:12mm) .. (O3.north east) node[pos=0.5, label=below:{{\large onr}}]{};
\draw[->,thick] (D0.south west) .. controls +(down:6mm) and +(down:6mm) .. (D3.south east) node[pos=0.6, label=below:{{\large dkr}}]{};

\draw[->,thick] (IO0.south west) .. controls +(down:12mm) and +(down:12mm) .. (IO3.south east) node[pos=0.5, label=below:{{\large onr}}]{};
\draw[->,thick] (IO0.south west) .. controls +(down:12mm) and +(3,-4) .. (D3.south east) node[pos=0.6, label={{\large ogr}}]{}; %(down:12mm)

\draw[->,thick] (O0.south west) .. controls (-4,-5)  and +(4,-4)  .. (D3.south east) node[pos=0.5, label=below:{{\large ogr}}]{}; %(down:40mm)
\draw[->,thick] (D0.north west) .. controls (-3,7) and +(up:12mm) .. (O3.north east) node[pos=0.5, label={{\large dlr}}]{};

\end{tikzpicture}
}
\caption{\label{fig:gnodeCov} Extended push-pull gossip model of individual gossip node with rounds of length 3 (i.e. $\Gm=3$) for the analysis of network coverage. Active states are red, passive ones blue.}
\end{center}
\end{figure}

Network coverage at time $t$ denotes the fraction of the gossip nodes that have seen the data element
at any point in time $t'$, with $t_0 \leq  t' \leq t$, where $t_0$ is the time the data element was introduced in the network. 
To analyse network coverage we extend the model of an individual node in Fig.~\ref{fig:gnodeCov} with four more states. These states are $I0$, $I1$, $I2$ and $I3$. A gossip node is in state $Ii$ if the data element is not in its cache and it has never seen the data element since it was introduced in the network. The latter is the case initially for most nodes, hence the name $I$: Initial O-state. If a node is in one of the other $O$-states this means that it does not have the data element in its cache currently, but it was in its cache at an earlier point in time, so the node has seen the data element since it was introduced for the first time.

The probability functions for the outgoing transitions of the $Ii$ nodes are the same as for their companion $O$-nodes. Also the probability functions of the incoming transitions, when they come from $I$-states, are the same. There are no incoming transitions from $D$-nodes of course, since passing by a $D$-node would mean that the data element has been in the cache of that node. Similarly, there is no transition from an $O$-state to an $I$-state.

The probability functions have to be updated slightly to take the two versions of the O-states into account.

$$
\begin{array}{l c l}
\mbox{\sf d\_loss\_reset}\;  m& = & (m_{O1}+m_{O2}+m_{O3})*P(OD|DO)*(\mbox{\sf noc })+\\ %e^{-2*(m_0+m_4+m_8)}+\\
 & &                                             (m_{I1}+m_{I2}+m_{I3})*P(OD|DO)*(\mbox{\sf noc } )+\\ %e^{-2*(m_0+m_4+m_8)}+\\
 & &                                             (m_{D1}+m_{D2}+m_{D3})*P(OD|DD)*(\mbox{\sf noc })\\[1em] %e^{-2*(m_0+m_4+m_8)}\\[1em]
\mbox{\sf d\_loss\_step}\;  m& = & (m_{O0}+m_{I0})*P(DO|OD)*(\mbox{\sf noc })+\\ %e^{-2*(m_0+m_4+m_8)}+\\
& &                                             m_{D0}*P(DO|DD)*(\mbox{\sf noc }) %e^{-2*(m_0+m_4)}
\end{array}
$$

and their dual probabilities, and

$$
\begin{array}{l c l}
\mbox{\sf o\_getd\_reset}\; m & = & (m_{D1}+m_{D2}+m_{D3})*(P(DO|OD) + P(DD|OD))*(\mbox{\sf noc })\\[1em] %e^{-2*(m_0+m_4+m_8)}\\[1em]
\mbox{\sf o\_getd\_step}\; m & = & m_{D0}*(P(OD|DO)+ P(DD|DO))*(\mbox{\sf noc }). %e^{-2*(m_0+m_4+m_8)}
\end{array}
$$

Similarly to the gossip model for replications we can define a 12 $\times$ 12 matrix $\vr{K}$ with indexes $i,j$ in $\{0,\cdots, 11\}$. The definition of the non-zero elements of this matrix can be found in the next section, as well as the additional set of difference equations that can be obtained from it.

\subsection{Transition matrix and difference equations for the gossip model in Fig.~\ref{fig:gnodeCov}}

Numbering the states in the model of Fig.~\ref{fig:gnodeCov} as O0=0, O1=1, O2=2, O3=3, D0=4, D1=5, D2=6, D3=7, I0=8, I1=9, I2=10 and I3=11, the non-empty elements of the $\vr{K}$ matrix are given by: 
$$
\begin{array}{l c l }
K_{0,3} & = & 1.0-(\mbox{\sf o\_getd\_reset}\; m)\\
K_{0,7} = K_{8,7} & = & \mbox{\sf o\_getd\_reset}\; m\\
K_{1,0}  =  K_{2,1}  =  K_{3,2} & = & 1.0 - (\mbox{\sf o\_getd\_step}\; m)\\
K_{1,4}  =  K_{2,5} = K_{3,6} & = & \mbox{\sf o\_getd\_step}\; m\\[1em]
K_{4,3} &=& \mbox{\sf d\_loss\_reset}\;  m\\
K_{4,7} &=& 1.0 - \mbox{\sf d\_loss\_reset}\;  m\\
K_{5,0} = K_{6,1} = K_{7,2} & = & \mbox{\sf d\_loss\_step}\;  m\\
K_{5,4} = K_{6,5} = K_{7,6} & = & 1.0 - \mbox{\sf d\_loss\_step}\;  m\\[1em]
K_{9,4} = K_{10,5} = K_{11,6} & = & \mbox{\sf o\_getd\_step}\; m\\
K_{9,8} = K_{10,9} = K_{11,10} &=&  1.0 - (\mbox{\sf o\_getd\_step}\; m)
\end{array}
$$

We also obtain the following set of additional four difference equations$^{\ref{note}}$ for the I-states of the model, representing the occupancy measure of state I0 by $m_{I0}$, I1 by $m_{I1}$, I2 by $m_{I2}$ and I3 by $m_{I3}$:

%$$
%\begin{array}{l c l}
%\mbox{IO0}\; m & = & m_9- m_9*(\mbox{\sf o\_getd\_step}\; m)\\ 
%\mbox{IO1}\; m & = & m_{10} - m_{10}*(\mbox{\sf o\_getd\_step}\; m)\\
%\mbox{IO2}\; m & = & m_{11} - m_{11}*(\mbox{\sf o\_getd\_step}\; m)\\
%\mbox{IO3}\; m & = & m_8 - m_8*(\mbox{\sf o\_getd\_reset}\; m)
%\end{array}
%$$

$$
\begin{array}{l c l}
m_{I0} & = & m_{I1}- m_{I2}*(\mbox{\sf o\_getd\_step}\; m)\\ 
m_{I1} & = & m_{I2} - m_{I2}*(\mbox{\sf o\_getd\_step}\; m)\\
m_{I2} & = & m_{I3} - m_{I3}*(\mbox{\sf o\_getd\_step}\; m)\\
m_{I3} & = & m_{I0} - m_{I0}*(\mbox{\sf o\_getd\_reset}\; m)
\end{array}
$$

\subsection{Transition matrix and difference equations for the gossip model in Fig.~\ref{fig:aggr_gnode_gm} (right)} 

Numbering the states in the model of Fig.~\ref{fig:aggr_gnode_gm} (right) as O=0, D=1, I=2 the non-empty elements of the $\vr{K}$ matrix are given by:

$$
\begin{array}{l c l }
K_{0,1} = K_{2,1} & = & \mbox{\sf get}\; m\\
K_{1,0} & = & \mbox{\sf loose}\; m\\
K_{0,0} = K_{2,2}  & = & 1.0 - (\mbox{\sf get}\; m)\\
K_{1,1} &=& 1.0 - (\mbox{\sf loose}\;  m)\\
\end{array}
$$

Representing the occupancy measure of state O by $m_O$, D by $m_D$, I by $m_I$ we obtain the following set of difference equations$^{\ref{note}}$ for the model in Fig.~\ref{fig:aggr_gnode_gm} (right):

$$
\begin{array}{l c l}
m_O & = & m_O - m_O*(\mbox{\sf get}\; m) + m_D*(\mbox{\sf loose}\;  m)\\
m_D & = & m_D + m_O*(\mbox{\sf get}\; m) - m_D*(\mbox{\sf loose}\;  m)\\
m_I & = & m_I - m_I*(\mbox{\sf get}\; m)
\end{array}
$$

\subsection{Transition matrix and difference equations for the gossip model in Fig.~\ref{fig:aggr_gnode_noloss_sync_6states}}

Numbering the states in the model of Fig.~\ref{fig:aggr_gnode_noloss_sync_6states} as O=0, D=1, I=2, FD=3, PD=4, LD=5, the non-empty elements of the $\vr{K}$ matrix are given by:

$$
\begin{array}{l c l }
K_{0,1} = K_{2,1} = K_{3,1} = K_{5,1}& = & \mbox{\sf get\_rep}\; m\\
K_{1,0} & = & \mbox{\sf loose\_rep}\; m\\
K_{0,0} = K_{2,2}  & = & 1.0 - (\mbox{\sf get\_rep}\; m) - (\mbox{\sf get\_exc}\; m)\\
K_{0,5} = K_{2,3} & = & \mbox{\sf get\_exc}\; m\\[1em]
K_{1,1} &=& 1.0 - (\mbox{\sf loose\_rep}\;  m)\\
K_{3,0} = K_{5,0} &=& \mbox{\sf loose\_exc}\;  m\\
K_{3,3} = K_{5,5}& = & 1.0 - (\mbox{\sf loose\_exc}\;  m) -  (\mbox{\sf get\_rep}\;  m)\\
K_{4,4} & = & 1.0\\[1em]
\end{array}
$$

Representing the occupancy measure of state O by $m_O$, D by $m_D$, I by $m_I$, FD by $m_{FD}$, PD by $m_{PD}$ and LD by $m_{LD}$, we also obtain the following set of difference equations$^{\ref{note}}$ for the model in Fig.~\ref{fig:aggr_gnode_noloss_sync_6states}:

$$
\begin{array}{l c l}
m_O & = & m_O - m_O*((\mbox{\sf get\_rep}\; m)+(\mbox{\sf get\_exc}\; m)) + m_D*(\mbox{\sf loose\_rep}\;  m)\\
& & + (m_{FD} + m_{LD})*(\mbox{\sf loose\_exc}\;  m)\\
m_D & = & (m_O+m_I+m_{FD}+m_{LD})*(\mbox{\sf get\_rep}\; m) + m_D - m_D*(\mbox{\sf loose\_rep}\;  m)\\
m_I & = & m_I - m_I*( \mbox{\sf get\_rep}\; m + \mbox{\sf get\_exc}\; m)\\
m_{FD} & = & m_{FD}- m_{FD}*(\mbox{\sf get\_rep}\;  m) + m_I*(\mbox{\sf get\_exc}\; m) - m_{FD}*(\mbox{\sf loose\_exc}\;  m)\\
m_{PD} & = & m_{PD}\\
m_{LD} & = & m_{LD}+ m_O*(\mbox{\sf get\_exc}\;  m) - m_{LD}*((\mbox{\sf loose\_exc}\;  m) + (\mbox{\sf get\_rep}\; m))
\end{array}
$$

\subsection{Detailed proofs}
\label{sec:proofs}

To prove: $\mathsf{dks}(\mu(t)) + \mathsf{dls}(\mu(t)) =1 \mbox{ for all }t.$. In the following $\frc(X)$ denotes the fraction of nodes in state $X$.

$\der \mathsf{dks}(\mu(t)) + \mathsf{dls}(\mu(t))
  \stp = \{\mbox{By Defs. of $\mathsf{dks}$ and $\mathsf{dls}$}\}\\[-1em]
  \stp {\phantom{=}} (\frc(O0)+\frc(D0))*(1-e^{-2*(\frc(O0)+\frc(D0))})+\\[-1em]
  \stp {\phantom{=}} (1-(\frc(O0)+\frc(D0)))+\\[-1em]
  \stp {\phantom{=}} \frc(O0)*(P(OD|OD)+ P(DD|OD))*e^{-2*(\frc(O0)+\frc(D0))}+\\[-1em]
  \stp {\phantom{=}} \frc(D0)*(P(OD|DD)+ P(DD|DD))*e^{-2*(\frc(O0)+\frc(D0))} +\\[-1em]
  \stp {\phantom{=}} \frc(O0)*P(DO|OD)*e^{-2*(\frc(O0)+\frc(D0))} +\\[-1em]
  \stp {\phantom{=}} \frc(D0)*P(DO|DD)*e^{-2*(\frc(O0)+\frc(D0))}\\[-1em]
  \stp = \{\mbox{Use that}\; \frc(D_i)(t) = \frac{\frc(D)(t)}{gmax+1} \mbox{and } \frc(O_i)(t)=\frac{\frc(O)(t)}{gmax+1}\}\\[-1em]
  \stp {\phantom{=}} (\frac{\frc(O)(t)}{gmax+1}+\frac{\frc(D)(t)}{gmax+1})*(1-*e^{-2/(gmax+1)})+\\[-1em]
  \stp {\phantom{=}} (1-(\frac{\frc(O)(t)}{gmax+1}+\frac{\frc(D)(t)}{gmax+1}))+\\[-1em]
  \stp {\phantom{=}} \frac{\frc(O)(t)}{gmax+1}*(P(OD|OD)+ P(DD|OD))*e^{-2/(gmax+1)}+\\[-1em]
  \stp {\phantom{=}} \frac{\frc(D)(t)}{gmax+1}*(P(OD|DD)+ P(DD|DD))*e^{-2/(gmax+1)} +\\[-1em]
  \stp {\phantom{=}} \frac{\frc(O)(t)}{gmax+1}*P(DO|OD)*e^{-2/(gmax+1)} +\\[-1em]
  \stp {\phantom{=}} \frac{\frc(D)(t)}{gmax+1}*P(DO|DD)*e^{-2/(gmax+1)}\\[-1em]
  \stp = \{\mbox{Simplify}\}\\[-1em]
  \stp {\phantom{=}} (\frac{\frc(O)(t)}{gmax+1}+\frac{\frc(D)(t)}{gmax+1})*(1-e^{-2*(\frc(O0)+\frac{\frc(D)(t)}{gmax+1})})+\\[-1em]
  \stp {\phantom{=}} (1-(\frac{\frc(O)(t)}{gmax+1}+\frac{\frc(D)(t)}{gmax+1}))+\\[-1em]
  \stp {\phantom{=}}  \frac{\frc(O)(t)}{gmax+1}*(P(OD|OD)+ P(DD|OD)+P(DO|OD))*e^{-2/(gmax+1)}+\\[-1em]
  \stp {\phantom{=}} \frac{\frc(D)(t)}{gmax+1}*(P(OD|DD)+ P(DD|DD)+P(DO|DD))*e^{-2/(gmax+1)}\\[-1em]
  \stp = \{\mbox{Use $P(OD|OD)+ P(DD|OD)+P(DO|OD)=1$ and}\\[-1em]
  \stp {\phantom{=}} \mbox{ $P(OD|DD)+ P(DD|DD)+P(DO|DD)=1$}\}\\[-1em]
  \stp {\phantom{=}} (\frac{\frc(O)(t)}{gmax+1}+\frac{\frc(D)(t)}{gmax+1})*(1-e^{-2*(\frc(O0)+\frac{\frc(D)(t)}{gmax+1})})+\\[-1em]
  \stp {\phantom{=}} (1-(\frac{\frc(O)(t)}{gmax+1}+\frac{\frc(D)(t)}{gmax+1}))+\\[-1em]
  \stp {\phantom{=}}  \frac{\frc(O)(t)}{gmax+1}*e^{-2/(gmax+1)}+\\[-1em]
  \stp {\phantom{=}} \frac{\frc(D)(t)}{gmax+1}*e^{-2/(gmax+1)}+\\[-1em]
  \stp = \{\mbox{Simplify}\}\\[-1em]
  \stp {\phantom{=}} 1
$\\[1em]

To prove: $P(OD|OD)+ P(DD|OD)+P(DO|OD)=1$\\[1em]

$\deriv P(OD|OD)+ P(DD|OD)+P(DO|OD)=1
  \stp = \{\mbox{Defs. of $P\_a'b'\_ab$}\}\\[-1em]
  \stp {\phantom{=}} \frac{c-s}{c} + \frac{s}{c}*(\frac{c-s}{n-s} + \frac{n-c}{n-s})\\[-1em]
  \stp = \{\mbox{Simplify}\}\\[-1em]
  \stp {\phantom{=}} \frac{c-s}{c} + \frac{s}{c}*\frac{n-s}{n-s} \\[-1em]
  \stp = \{\mbox{Simplify}\}\\[-1em]
   \stp {\phantom{=}} \frac{c-s}{c} + \frac{s}{c}\\[-1em]
   \stp = \{\mbox{Simplify}\}\\[-1em]
   \stp {\phantom{=}} 1
$ \\[1em]

The proof for $P(OD|DD)+ P(DD|DD)+P(DO|DD)=1$ is very similar.

\end{document}